\documentclass[11pt,a4paper]{article} 
\usepackage{bm}
\usepackage{color}
\usepackage{amsmath}
\usepackage{amsthm}
\usepackage{amssymb}
\usepackage{graphicx}
\usepackage{verbatim}
\usepackage[breaklinks]{hyperref}

\newtheorem{lemma}{Lemma}
\newtheorem{theorem}{Theorem}

\newtheorem{claim}{Claim}
\theoremstyle{definition}
\newtheorem{definition}{Definition}

\definecolor{aocolour}{rgb}{0.7,0.8,1}
\definecolor{omcolour}{rgb}{1,0.4,0.4}

\newcommand{\set}[2]{\ensuremath{ \{ \, #1 \mid #2 \, \} }} 
\newcommand{\setbig}[2]{\big\{ \: #1 \;\big|\; #2 \: \big\}}
\renewcommand{\emptyset}{\varnothing}
\renewcommand{\epsilon}{\varepsilon}

\newcommand{\forward}[1]{\overrightarrow{#1}}
\newcommand{\reverse}[1]{[#1]}
\newcommand{\forwardQ}{\overrightarrow{Q}}
\newcommand{\reverseQ}{[Q]}
\newcommand{\revrevQ}{[[Q]]}
\newcommand{\revforwardQ}{[\overrightarrow{Q}]}

\newcommand{\m}{m}
\newcommand{\mi}[1]{m_{#1}}
\newcommand{\me}{m_{\mathrm{e}}}
\newcommand{\ma}{m_a}

\newcommand{\seepage}[2][see]{\marginpar{\scriptsize (\text{#1} p.~\pageref{#2})}}

\begin{document}

\title{State complexity of halting, returning and reversible graph-walking automata}

\author{Olga Martynova\thanks{%
	Department of Mathematics and Computer Science,
	St.~Petersburg State University, Russia,
	e-mail: \texttt{olga22mart@gmail.com}.}
	\and
	Alexander Okhotin\thanks{%
	Department of Mathematics and Computer Science,
	St.~Petersburg State University, Russia,
	e-mail: \texttt{alexander.okhotin@spbu.ru}.}
}

\maketitle

\begin{abstract}
Graph-walking automata (GWA) traverse graphs by moving between the nodes following the edges,
using a finite-state control to decide where to go next.
It is known
that every GWA can be transformed to a GWA that halts on every input,
to a GWA returning to the initial node in order to accept,
and to a reversible GWA.
This paper establishes lower bounds on the state blow-up of these transformations,
as well as closely matching upper bounds.
It is shown that making an $n$-state GWA traversing $k$-ary graphs
halt on every input
requires at most $2nk+1$ states and at least $2(n-1)(k-3)$ states in the worst case;
making a GWA return to the initial node before acceptance
takes at most $2nk+n$ and at least $2(n-1)(k-3)$ states in the worst case;
Automata satisfying both properties at once
have at most $4nk+1$ and at least $4(n-1)(k-3)$ states in the worst case.
Reversible automata have at most $4nk+1$ and at least $4(n-1)(k-3)-1$ states in the worst case.
\\

\end{abstract}

\tableofcontents


\sloppy

\section{Introduction}

Graph-walking automata (GWA) are finite automata
that traverse labelled undirected graphs.

On the one hand, this is a model of a robot with limited memory
navigating a discrete environment.
There is an early result by Budach~\cite{Budach}
that for every automaton there is a graph that it cannot fully explore;
a short proof of this fact was later given by Fraigniaud et al.~\cite{FraigniaudIlcinkasPeerPelcPeleg}.
This work has influenced
the current research on algorithms for graph traversal
using various small-memory models,
equipped with a limited number of pebbles, etc.~\cite{DisserHackfeldKlimm,ElmasryHagerupKammer}.

On the other hand, GWA
naturally generalize such important models
as tree-walking automata~\cite{BojanczykColcombet_det} (TWA)
and two-way finite automata (2DFA).
More generally, a GWA can represent various models of computation,
if a graph is regarded as the space of memory configurations,
and every edge accordingly represents an operation on the memory.
This way, quite a few models in automata theory and in complexity theory,
such as multi-head and multi-tape automata and space-bounded complexity classes,
can be regarded as GWA.
Then, some results on GWA apply to all these models.

Among such results, there are transformations of GWA
to several important subclasses:
to automata that halt on every input graph;
to automata that return to the initial node in order to accept;
to reversible automata.
Such transformations have earlier been established for various automaton models,
using a general method discovered by Sipser~\cite{Sipser_halting},
who constructed a halting 2DFA
that traverses the tree of computations of a given 2DFA
leading to an accepting configuration,
in search of an initial configuration.
Later, Kondacs and Watrous~\cite{KondacsWatrous}
ensured the reversibility and optimized this construction for the number of states,
motivated by the study of quantum automata.
Sipser's idea has been adapted
to proving that reversible space equals deterministic space~\cite{LangeMcKenzieTapp},
to making tree-walking automata halt~\cite{MuschollSamuelidesSegoufin},
to complementing 2DFA~\cite{GeffertMereghettiPighizzini},
to making multi-head automata reversible~\cite{Morita},
etc.
Each transformation leads to a certain blow-up in the number of states,
usually between linear and quadratic.
No lower bounds on the transformation to halting
have been established yet.
For the transformation to reversible,
a lower bound exists for the case of 2DFA~\cite{rev2dfa},
but it is quite far from the known upper bound.

For the general case of GWA,
constructions of halting, returning and reversible automata
were given by Kunc and Okhotin~\cite{KuncOkhotin_reversible},
who showed that an $n$-state GWA
operating on graphs with $k$ edge labels
can be transformed to a returning GWA with $3nk$ states
and to a reversible GWA with $6nk+1$ states, which is always halting.
Applied to special cases of GWA, such as TWA or multi-head automata,
these generic constructions produce fewer states
than the earlier specialized constructions.

The goal of this paper is to obtain lower bounds
on the complexity of these transformations.
To begin with,
the constructions by Kunc and Okhotin~\cite{KuncOkhotin_reversible}
are revisited in Section~\ref{section_upper_bounds},
and it turns out that the elements they are built of
can be recombined more efficiently,
resulting in improved upper bounds based on the existing methods.
This way, the transformation to a returning GWA is improved to use $2nk+n$ states,
the transformation to halting can use $2nk+1$ states,
and constructing a reversible GWA (which is both returning and halting)
requires at most $4nk+1$ states.
The main result of the paper is that, with these improvements,
each of these constructions is asymptotically optimal.

The lower bounds are proved according to the following plan.
For each $n$ and $k$,
one should construct an $n$-state automaton
operating on graphs with $k$ direction labels,
so that any returning, halting or reversible automaton recognizing the same language
would require many states.
The $n$-state automaton follows a particular path in an input graph
in search for a special node.
The node is always on that path, so that the automaton naturally encounters it if it exists.
On the other hand, the graph is constructed,
so that getting back is more challenging.

The graph is made of elements called \emph{diodes},
which are easy to traverse in one direction
and hard to traverse backwards.
Diodes are defined in Section~\ref{section_diode},
where it is shown that a GWA needs to employ extra states
to traverse a diode backwards.

The graph used in all lower bound arguments,
constructed in Section~\ref{section_returning},
has a main path made of diodes leading to a special node,
which makes returning more complicated,
so that a returning automaton needs at least $2(n-1)(k-3)$ states.
A variant of this graph containing a cycle made of diodes,
presented in Section~\ref{section_halting},
poses a challenge to a halting automaton,
which needs at least $2(n-1)(k-3)$ states.
Section~\ref{section_returning_and_halting} combines the two arguments
to establish a lower bound of $4(n-1)(k-3)$
on the number of states of an automaton
that is returning and halting at the same time.
This bound is adapted to reversible automata in Section~\ref{section_reversible}:
at least $4(n-1)(k-3)-1$ states are required.

Overall, each transformation requires ca.~$C \cdot nk$ states in the worst case,
for a constant $C$.
Each transformation has its own constant $C$,
and these constants are determined precisely.

\section{Graph-walking automata and their subclasses}

The definition of graph-walking automata (GWA)
is an intuitive extension of two-way finite automata (2DFA)
and tree-walking automata (TWA).
However, formalizing it requires extensive notation.
First, there is a notion of a \emph{signature},
which is a generalization of an alphabet to the case of graphs.

\begin{definition}[Kunc and Okhotin~\cite{KuncOkhotin_reversible}]
\emph{A signature $S$} consists of
\begin{itemize}
\item A finite set $D$ of directions, that is, labels attached to edge end-points;
\item A 
	bijection
	$- \colon D \to D$ providing an opposite direction,
	with $-(-d) = d$ for all $d \in D$;
\item A finite set $\Sigma$ of node labels;
\item A non-empty subset $\Sigma_0 \subseteq \Sigma$ of possible labels of the initial node;
\item A set of directions $D_a \subseteq D$ for every label $a \in \Sigma$. 
	Every node labelled with $a$ must be of degree $|D_a|$, 
	with the incident edges corresponding to the elements of $D_a$.
\end{itemize}
\end{definition}

Graphs are defined over a signature,
like strings over an alphabet.

\begin{definition}
\emph{A graph over a signature $S = (D, -, \Sigma, \Sigma_0, (D_a)_{a \in \Sigma})$}
is a quadruple $(V, v_0, +, \lambda)$, where
\begin{itemize}
\item $V$ is a finite set of nodes;
\item $v_0 \in V$ is the initial node;
\item $+ \colon V \times D \to V$ is a partial function, such that
	if $v+d$ is defined, then $(v+d) + (-d)$ is defined and equals $v$;
\item a total mapping $\lambda \colon V \to \Sigma$, such that
	$v+d$ is defined
	if and only if
	$d \in {D_{\lambda(v)}}$,
	and
%
	$\lambda(v) \in \Sigma_0$
	if and only if
	$v = v_0$.
\end{itemize}
\end{definition}


Once graphs are formally defined,
a graph-walking automaton is defined similarly to a 2DFA.

\begin{definition}
\emph{A (deterministic) graph-walking automaton (GWA) over a signature 
$S = (D, -, \Sigma, \Sigma_0, (D_a)_{a \in \Sigma})$} 
is a quadruple $A = (Q, q_0, F, \delta)$, where
\begin{itemize}
\item $Q$ is a finite set of states;
\item $q_0 \in Q$ is the initial state;
\item $F \subseteq Q \times \Sigma$ is a set of acceptance conditions;
\item $\delta \colon (Q \times \Sigma) \setminus F \to Q \times D$ is 
a partial transition function, with $\delta(q, a) \in Q \times D_a$ for all $a$ 
and $q$ where $\delta$ is defined.
\end{itemize}
A computation of a GWA on a graph $(V, v_0, +, \lambda)$
is a uniquely defined
sequence of configurations $(q, v)$, with $q \in Q$ and $v \in V$.
It begins with $(q_0, v_0)$
and proceeds from $(q, v)$ to $(q', v+d)$, where $\delta(q, \lambda(v))=(q', d)$.
The automaton accepts by reaching $(q, v)$ with $(q, \lambda(v)) \in F$.
\end{definition}

On each input graph, a GWA can accept, reject or loop.
There is a natural subclass of GWA that never loop.

\begin{definition}
A graph-walking automaton is said to be halting,
if its computation on every input graph is finite.
\end{definition}

Another property is getting back to the initial node before acceptance:
if a GWA is regarded as a robot,
it returns to its hangar,
and for a generic model of computation,
this property means cleaning up the memory.

\begin{definition}
A graph-walking automaton $A = (Q, q_0, F, \delta)$ 
over a signature $S = (D, -, \Sigma, \Sigma_0, (D_a)_{a \in \Sigma})$ is called returning, 
if $F \subseteq Q \times \Sigma_0$, which means that
it can accept only in the initial node.
\end{definition}

A returning automaton is free to reject in any node, and it may also loop,
that is, it need not be halting.

The next, more sophisticated property is \emph{reversibility},
meaning that, for every configuration,
the configuration at the previous step
can be uniquely reconstructed.
This property is essential in quantum computing,
whereas irreversibility in classical computers
causes energy dissipation,
which is known as \emph{Landauer's principle}~\cite{Landauer}.

The definition of reversibility begins with the property
that every state is reachable from only one direction.

\begin{definition}
A graph-walking automaton $A = (Q, q_0, F, \delta)$ over a signature 
$S = (D, -, \Sigma, \Sigma_0, (D_a)_{a \in \Sigma})$ is called direction-determinate, 
if there is a function $d \colon Q \to D$,
such that, for all $p \in Q$ and $a \in \Sigma$,
if $\delta(p, a)$ is defined,
then $\delta(p,a) = (q,d(q))$ for some $q \in Q$.
%
\end{definition}

\begin{definition}
A graph-walking automaton $A = (Q, q_0, F, \delta)$ 
over a signature $S = (D, -, \Sigma, \Sigma_0, (D_a)_{a \in \Sigma})$ is called reversible, if
\begin{itemize}
\item $A$ is direction-determinate;
\item
for all $a \in \Sigma$ and $q \in Q$,
there is at most one state $p$,
such that $\delta(p, a)=(q, d(q))$;
in other words,
knowing a state and a previous label, one can determine the previous state;
\item The automaton is returning, and for each $a_0 \in \Sigma_0$ there 
exists at most one such state $q$, that $(q,a_0) \in F$.
\end{itemize}
\end{definition}

In theory, a reversible automaton may loop, but only through the initial configuration.
In this case, it can be made halting by introducing an extra initial state.

Every GWA can be transformed to each of the above subclasses~\cite{KuncOkhotin_reversible}.
In the next section, the known transformations
will be explained and slightly improved.

\section{Upper bounds revisited}\label{section_upper_bounds}

Before establishing the lower bounds on all transformations,
the existing constructions of Kunc and Okhotin~\cite{KuncOkhotin_reversible}
will be somewhat improved by using fewer states.
This is achieved by recombining the elements of the original construction,
and, with these improvements,
the constructions shall be proved asymptotically optimal.

The basic element for constructing reversible automata
is a lemma by Kunc and Okhotin~\cite{KuncOkhotin_reversible}
(Lemma~\ref{reversibility_construction_lemma} below),
which implements Sipser's idea of backtracking the tree of computations
coming to an accepting configuration
in the general case of graph-walking automata.
The lemma assumes that the automaton is already direction-determinate,
and it makes a further technical assumption
that whenever the automaton has any behaviour defined on a pair $(q, a)$---transition or acceptance---%
the label $a$ must support incoming transitions in the direction $d(q)$.

\begin{definition}[{Kunc and Okhotin~\cite[Defn.~6]{KuncOkhotin_reversible}}]
A direction-determinate automaton $A = (Q, q_0, \delta, F)$
is \emph{without inaccessible transitions},
if, for all pairs $(q, a) \in (Q \times \Sigma)$ with $(q, a) \notin \{q_0\} \times \Sigma_0$,
if $\delta_a(q)$ is defined or $(q,a) \in F$,
then $-d(q) \in D_a$.
\end{definition}

If there are any such transitions,
they can be removed without affecting any computations.

\begin{lemma}[{Kunc and Okhotin~\cite[Lemma~6]{KuncOkhotin_reversible}}]\label{reversibility_construction_lemma}
For every direction-determinate automaton $\mathcal{A} = (Q, q_0, \delta, F)$
without inaccessible transitions,
there exists a reversible automaton
$\mathcal{B} = (\forwardQ \cup \reverseQ, \delta', F')$
without an initial state,
and with states $\forwardQ = \set{\forward{q}}{q \in Q}$
and $\reverseQ = \set{\reverse{q}}{q \in Q}$,
which simulates $\mathcal{A}$ as follows.
Each state $q \in Q$ has two corresponding states in $\mathcal{B}$:
a \emph{forward state} $\forward{q}$
accessible in the same direction $d'(\forward{q}) = d(q)$,
and a \emph{backward state} $\reverse{q}$
accessible in the opposite direction $d'(\reverse{q}) = -d(q)$.
The acceptance conditions of $\mathcal{B}$ are
$F'=\setbig{(\reverse{\delta_{a_0}(q_0)}, a_0)}
{a_0 \in \Sigma_0,\ \delta_{a_0}(q_0) \text{ is defined}}$.
For every finite graph $(V, v_0, \lambda, +)$, its node $\widehat{v} \in V$
and a state $\widehat{q} \in Q$ of the original automaton,
for which 
$(\widehat{q}, \lambda(\widehat{v})) \in F$
and 
$-d(\widehat{q}) \in D_{\lambda(\widehat{v})}$,
the computation of $\mathcal{B}$
beginning in the configuration
$(\reverse{\widehat{q}}, \widehat{v}-d(\widehat{q}))$
has one of the following two outcomes.
\begin{itemize}
\item
	If $\mathcal{A}$ accepts this graph
	in the configuration $(\widehat{q}, \widehat{v})$,
	and if $(\widehat{q}, \widehat{v}) \neq (q_0,v_0)$,
	then $\mathcal{B}$ accepts in the configuration
	$(\reverse{\delta_{\lambda(v_0)}(q_0)}, v_0)$.
\item
	Otherwise, $\mathcal{B}$
	rejects in $(\forward{\widehat{q}}, \widehat{v})$.
\end{itemize}
\end{lemma}

In particular, $A$ in configuration $(q, v)$
is simulated by $B$ forward in the configuration $(\forward{q}, v)$
and backward in the configuration $(\reverse{q}, v-d(q))$.
Note that the latter configuration is shifted by one node along the computation.

Note that the computation of $A$ starting from the initial configuration
can reach at most one accepting configuration $(q, v)$,
whereas for any other accepting configuration $(q, v)$,
the automaton $B$ will not find the initial configuration
and will reject as stated in the lemma.

To transform a given $n$-state GWA $\widehat{A}$
over a signature with $k$ directions to a returning automaton,
Kunc and Okhotin~\cite{KuncOkhotin_reversible}
first transform it to a direction-determinate automaton $A$ with $nk$ states;
let $B$ be the $2nk$-state automaton
obtained from $A$ by Lemma~\ref{reversibility_construction_lemma}.
Then they construct an automaton that first operates as $A$,
and then, after reaching an accepting configuration,
works as $B$ to return to the initial node.
This results in a returning direction-determinate automaton with $3nk$ states.

If the goal is just to return, and remembering the direction is not necessary,
then $2nk+n$ states are actually enough.

\begin{theorem}\label{gwa_to_returning_transformation_theorem}
For every $n$-state GWA
over a signature with $k$ directions,
there exists a returning automaton
with $2nk+n$ states recognizing the same set of graphs.
\end{theorem}

Indeed, the original automaton $\widehat{A}$ can be first simulated as it is,
and once it reaches an accepting configuration,
one can use the same automaton $B$
as in the original construction
to return to the initial node.
There is a small complication in the transition from $\widehat{A}$ to $B$,
because in the accepting configuration,
the direction last used is unknown.
This is handled by cycling
through all possible previous configurations of $A$
at this last step,
and executing $B$ from each of them.
If the direction is guessed correctly, then $B$ finds the initial configuration and accepts.
Otherwise, if the direction is wrongly chosen, $B$ returns back,
and then, instead of rejecting,
it is executed again starting from the next direction.
One of these directions
leads it back to the initial node.

\begin{proof}[Proof of Theorem~\ref{gwa_to_returning_transformation_theorem}]
Let $\widehat{A} = (Q, q_0, \delta, F)$ be the given automaton.
Let $A$ be a direction-determinate automaton
with the set of states $Q \times D$
that recognizes the same set of graphs~\cite[Lemma~1]{KuncOkhotin_reversible}.
Assume that all inaccessible transitions are removed from $A$.

Let $B$ be the automaton
obtained from $A$ by Lemma~\ref{reversibility_construction_lemma}:
this is a reversible automaton without initial state,
and it uses the set of states $\forwardQ \cup \reverseQ$,
where $\forwardQ = \set{\forward{(q, d)}}{q \in Q, d \in D}$
and $\reverseQ = \set{\reverse{(q, d)}}{q \in Q, d \in D}$.
Its transition function is denoted by $\delta'$.
There are $2nk$ states in $B$.

Assume any linear order on $D$.

A new automaton $C$ is constructed
with the set of states $Q \cup \forwardQ \cup \reverseQ$ containing $n+2nk$ states,
and with a transition function $\delta''$, to be defined below.
If the initial configuration of $C$
is an accepting configuration of $\widehat{A}$,
then $C$ immediately accepts as well.
Otherwise, $C$ begins by simulating $\widehat{A}$ in the states $Q$.
\begin{align*}
	\delta''(q, a) &= \delta(q, a),
		&& \text{for } q \in Q \text{ and } a \in \Sigma,
		\text{ if } \delta(q, a) \text{ is defined}
\intertext{%
If the simulated $\widehat{A}$ reaches an accepting configuration $(\widetilde{q}, \widehat{v})$,
this means that $A$, operating on the same graph, would reach an accepting configuration of the form
$((\widetilde{q}, \widehat{d}), \widehat{v})$, for some direction $\widehat{d} \in -D_a$.
However, $C$ does not know this direction $\widehat{d}$,
so it tries moving in all directions in $D_a$, beginning with the least one.
}
	\delta''(\widetilde{q}, a) &= (\reverse{(\widetilde{q}, -\min D_a)}, \min D_a)
		&& \text{for all $(\widetilde{q}, a) \in F \setminus (\{q_0\} \times \Sigma_0)$}
\intertext{%
Let $d=-\min D_a$.
In the notation of Lemma~\ref{reversibility_construction_lemma},
$\widehat{q}=(\widetilde{q}, \widehat{d})$
and $d(\widehat{q}) = \widehat{d}$
and $-\widehat{d} \in D_a$.
According to the lemma, if the direction was correctly guessed as $d=\widehat{d}$,
then $B$, having started in the configuration $(\reverse{\widehat{q}}, \widehat{v}-d)$,
accepts at the initial node.
The automaton $C$ simulates $B$ to do the same.
}
	\delta''(\reverse{(q, d)}, a) &= \delta'(\reverse{(q, d)}, a),
		&& \text{for all } a \in \Sigma \text{ and } \reverse{(q, d)} \in \reverseQ
			\\
	\delta''(\forward{(q, d)}, a) &= \delta'(\forward{(q, d)}, a),
		&& \text{for all } a \in \Sigma \text{ and } \forward{(q, d)} \in \forwardQ
\intertext{%
If the direction was wrongly guessed as $d \neq \widehat{d}$,
then $(\widetilde{q}, d)$ is still a valid state of $A$,
so, by the lemma,
$B$ returns to the configuration
$(\forward{(\widetilde{q}, d)}, \widehat{v})$,
in which it would reject.
The automaton $C$, instead of rejecting,
tries the next available direction from $D_a$.
}
	\delta''(\forward{(\widetilde{q}, d)}, a) &=
	(\reverse{(\widetilde{q}, d')}, -d')
		&& \text{for all $(\widetilde{q}, a) \in F \setminus (\{q_0\} \times \Sigma_0)$}, \:
		d'=-next_{-d}(D_a)
\end{align*}
Here $next_{-d}(D_a)$ denotes the least element of $D_a$ greater than $-d$.
The above transition is defined, assuming that $-d$ is not the greatest element of $D_a$.
Since one of the directions in $D_a$ is the true $-\widehat{d}$,
it is eventually found, and the case of all directions failing need not be considered.

The acceptance conditions of $C$ are the same as in $B$,
and so $C$ is returning.
As argued above, $C$ recognizes the same set of graphs as $\widehat{A}$.
\end{proof}

Kunc and Okhotin~\cite{KuncOkhotin_reversible}
did not consider halting automata separately.
Instead, they first transform an $n$-state GWA
to a $3nk$-state returning direction-determinate automaton,
then use Lemma~\ref{reversibility_construction_lemma}
to obtain a $6nk$-state reversible automaton,
and add an extra initial state to start it.
The resulting $(6nk+1)$-state automaton is always halting.

If only the halting property is needed,
then the number of states can be reduced. 

\begin{theorem}\label{dirdet_to_halting_transformation_theorem}
For every $n$-state direction-determinate automaton,
there exists a $(2n+1)$-state halting and direction-determinate automaton
that recognizes the same set of graphs.
\end{theorem}

First, an $n$-state automaton $\widehat{A}$
is transformed to a direction-determinate $nk$-state automaton $A$,
and Lemma~\ref{reversibility_construction_lemma}
is used to construct a $2nk$-state automaton $B$.
Then, the automaton $B$ is \emph{reversed}
by the method of Kunc and Okhotin~\cite{KuncOkhotin_reversible},
resulting in an automaton $B^R$ with $2nk+1$ states
that carries out the computation of $B$ backwards.
The automaton $B^R$ is a halting automaton that recognizes the same set of graphs as $\widehat{A}$:
it starts in the initial configuration,
and if $B$ accepts from an accepting configuration of $A$,
then $B^R$ finds this configuration and accepts;
otherwise, $B^R$ halts and rejects.

\begin{proof}[Proof of Theorem~\ref{dirdet_to_halting_transformation_theorem}.]
Let $A$ be the original automaton, let $Q$ be its set of states.
Assume that all inaccessible transitions have already been removed from $A$.
Lemma~\ref{reversibility_construction_lemma}
is used to construct a $2n$-state reversible automaton $B$ without an initial state.
The latter automaton $B$ is then subjected to another transformation:
by the method of Kunc and Okhotin~\cite{KuncOkhotin_reversible},
$B$ is transformed to a \emph{reversed} automaton $B^R$,
which carries out the computation of $B$ backwards,
starting from the accepting configuration of $B$,
and using a state $\reverse{\bm{q}}$
to simulate $B$ in a state $\bm{q}$;
in the simulation, the configurations are shifted by one node relative to $B$,
so that $B$ in $(\bm{q}, v)$ corresponds to $B^R$ in $(\bm{q}, v-d(\bm{q}))$.
Since $B$ has no initial configuration defined,
the acceptance conditions of $B^R$ are not defined either,
and shall be supplemented in the following.
The automaton $B^R$ has $2n+1$ states.

By Lemma~\ref{reversibility_construction_lemma},
the set of states of $B$ is $Q'=\reverseQ \cup \forwardQ$.
Then, $B^R$ has the set of states
	$Q'' = \revrevQ \cup \revforwardQ \cup \{q''_0\}$,
where $\revrevQ=\set{\reverse{\reverse{q}}}{q \in Q}$
and $\revforwardQ=\set{\reverse{\forward{q}}}{q \in Q}$,
where $q''_0$ is the new initial state.
The directions of states in $B$ are
$d'(\reverse{q})=-d(q)$ and $d'(\forward{q})=d(q)$,
and $B^R$ uses the opposite directions:
$d''(\reverse{\reverse{q}}) = d(q)$ and
$d''(\reverse{\forward{q}}) = -d(q)$,
for all $q \in Q$.

Let $\delta$ be the transition function of $A$,
let the transition function of $B$ be $\delta'$,
and let $B^R$ use $\delta''$.


The transitions of $B^R$ are defined by joining two constructions
by Kunc and Okhotin~\cite{KuncOkhotin_reversible},
and are listed below.
Transitions in the initial state $q''_0$ correspond to accepting conditions of $B$.
\begin{equation*}
	\delta''_{a_0}(q''_0) = \begin{cases}
	\reverse{\reverse{\delta_{a_0}(q_0)}},
		& \text{if $\delta_{a_0}(q_0)$ is defined in $A$} \\
	\text{undefined},
		& \text{otherwise}
	\end{cases}
\end{equation*}
The rest of the transitions carry out the computation backwards;
each of them is uniquely defined, because $B$ is reversible.
\begin{equation*}
	\delta''_a(\reverse{\bm{q}}) = \begin{cases}
	\reverse{\bm{p}},
		& \text{if $\delta'_a(\bm{p})=\bm{q}$} \\
	\text{undefined},
		& \text{otherwise}
	\end{cases}
\end{equation*}
The acceptance conditions of $B^R$ are defined,
so that it accepts a graph if so does $A$.
\begin{equation*}
	F''=\setbig{(\reverse{\reverse{q}}, a)}{(q, a) \in F}
	\cup \setbig{(q''_0, a_0)}{(q_0, a_0) \in F}
\end{equation*}
Overall, after being shifted by one node \emph{twice},
the state $\reverse{\reverse{q}}$ corresponds to the state $q$ of $A$,
and $B^R$ in a configuration $(\reverse{\reverse{q}}, v)$
corresponds to $A$ in $(q, v)$.

It is claimed that the automaton $B^R$ is a halting automaton
that recognizes the same set of graphs as $A$.

Its transitions are reversible by construction,
and therefore it can loop only through the initial configuration.
But since its initial state is not reenterable,
it cannot loop at all, and is therefore halting.

To see that $B^R$ recognizes the same graphs as $A$,
first assume that $A$ accepts some graph in a configuration $(\widehat{q}, \widehat{v})$.
Then, $B$ backtracks from $(\reverse{\widehat{q}}, \widehat{v}-d(\widehat{q}))$
to the configuration $(\reverse{\delta_{a_0}(q_0)}, v_0)$.
The automaton $B^R$ in turn moves from the configuration $(q''_0, v_0)$
to 
$(\reverse{\reverse{\widehat{q}}}, \widehat{v})$
and accepts there.

Conversely, if $B^R$ accepts a graph,
then it does so in a configuration $(\reverse{\reverse{\widehat{q}}}, \widehat{v})$.
The corresponding computation of $B$ proceeds from 
$(\reverse{\widehat{q}}, \widehat{v}-d(\widehat{q}))$
to  $(\reverse{\delta_{a_0}(q_0)}, v_0)$,
and then, by Lemma~\ref{reversibility_construction_lemma},
$A$ accepts this graph in the configuration $(\widehat{q}, \widehat{v})$.
\end{proof}

The construction of a reversible automaton with $6nk+1$ states
can be improved to $4nk+1$ by merging the automata $B$ and $B^R$.
The new automaton first works as $B^R$
to find the accepting configuration of $A$.
If it finds it, then it continues as $B$ to return to the initial node.
In addition, this automaton halts on every input.

\begin{theorem}\label{dirdet_to_reversible_transformation_theorem}
For every $n$-state direction-determinate automaton
there exists a $(4n+1)$-state reversible and halting automaton
recognizing the same set of graphs.
\end{theorem}
\begin{proof}
%
Let $A$ be a given direction-determinate automaton with $n$ states.
Assume that it is without \emph{inaccessible transitions}.


First, as in the proof of Theorem~\ref{dirdet_to_halting_transformation_theorem},
let $B$ be the reversible automaton without an initial state
constructed by Lemma~\ref{reversibility_construction_lemma},
and then let $B^R$ be another automaton with reversible transitions
that simulates $B$ backwards.


The goal is to combine $B^R$ with $B$
to obtain a new reversible and halting automaton $C$.
The automaton $C$ first operates as $B^R$,
which is also known to be halting.
Then, if $A$ rejects or loops, then $B^R$ halts and rejects, and $C$ rejects accordingly.
If $A$ accepts immediately in its initial configuration, then so do $B^R$ and $C$.

Assume that $A$ accepts in a non-initial configuration $(\widehat{q}, \widehat{v})$.
Then, $B^R$ arrives at a configuration
$(\reverse{\reverse{\widehat{q}}}, \widehat{v})$ and accepts.
The automaton $C$ needs to return to the initial node,
so it does not accept immediately, as does $B^R$,
and instead enters the configuration
$(\reverse{\widehat{q}}, \widehat{v} - d(\widehat{q}))$ 
and continues as $B$. 
By definition, $B$, having started from this configuration,
reaches the configuration $(\reverse{\delta_{a_0}(q_0)}, v_0)$ and accepts there;
$C$ does the same.

The set of states of $C$ is the union of the sets of states of $B$ and of $B^R$.
There are $4n+1$ states in total,
and they are enterable in the same directions as in $B$ and in $B^R$.
The initial state of $C$ is $q''_0$ from $B^R$.
Its transitions are all the transitions of $B$ and $B^R$,
as well as the following transitions that transfer control from $B^R$ to $B$.
These transitions are defined
for every accepting pair $(\reverse{\reverse{\widehat{q}}}, a)$ of $B^R$,
unless it is initial.
\begin{align*}
	\delta'''_a(\reverse{\reverse{\widehat{q}}}) = \reverse{\widehat{q}}
	&& ((\widehat{q}, a) \in F \setminus \{q_0\} \times \Sigma_0)
\end{align*}
The acceptance conditions of $C$ are the same as in $B$.

By the construction, the automaton $C$
accepts the same set of graphs as $A$.
It has $4n+1$ states.
It should be proved that it is reversible.

For every initial label, $C$ has at most one acceptance condition involving this label,
since this is the case for $B$.
The automaton $C$ is returning.
It remains to prove that, for every label $a$ and for every state $\bm{q}$ of $C$, 
there is at most one state $\bm{p}$ with $\delta'''_a(\bm{p}) = \bm{q}$.

The automata $B$ and $B^R$ had this property by construction.
The only new transitions are transitions of the form
$\delta'''_a(\reverse{\reverse{\widehat{q}}}) = \reverse{\widehat{q}}$,
for non-initial acceptance conditions $(\widehat{q}, a)$ of the automaton $A$.
It is enough to show that $B$ does not have any other way of reaching a state $\reverse{\widehat{q}}$
from the label $a$.

Some details of the construction of $B$ need to be examined~\cite[Lemma~6]{KuncOkhotin_reversible}.
There are two kinds of states in $B$: $\reverse{p}$ and $\forward{p}$.
By the construction of $B$, the state $\reverse{\widehat{q}}$
can potentially be reached by the following two transitions.

First, this could be a transition of the form $\delta'_a(\reverse{p}) = \reverse{\widehat{q}}$
~\cite[Eq.~(1)]{KuncOkhotin_reversible}.
However, by the definition of $B$, this transition is defined
only if $A$ has a transition $\delta_a(\widehat{q}) = p$.
This is impossible, since $(\widehat{q}, a)$ is an accepting pair for $A$.

The other possible transition is $\delta'_a(\forward{p}) = \reverse{\widehat{q}}$
~\cite[Eq.~(3)]{KuncOkhotin_reversible}.
For this transition to be defined,
$A$ must have a transition $\delta_a(\widehat{q}) = \delta_a(p)$,
and $\delta_a(p)$ must accordingly be defined.
Again, this cannot be the case,
because the pair $(\widehat{q}, a)$ is accepting in $A$.

The definition of $B$ includes one further type of transitions
leading to a state $\reverse{\widehat{q}}$
~\cite[Eq.~(5)]{KuncOkhotin_reversible}.
Those transitions are actually never used and can be omitted;
they were defined in the original paper for the sole reason
of making the transition function bijective.

This confirms that the transfer of control from $B^R$ to $B$
is done reversibly,
and so $C$ is reversible.
\end{proof}

With the upper bounds improved,
it is time to establish asymptotically matching lower bounds.

\section{Construction of a ``diode''}\label{section_diode}

Lower bounds on the size of GWA obtained in this paper
rely on quite involved constructions of graphs
that are easy to traverse from the initial node to the moment of acceptance,
whereas traversing the same path backwards is hard.
An essential element of this construction
is a subgraph called a \emph{diode};
graphs in the lower bound proofs are made of such elements.

A diode is designed to replace an $(a, -a)$-edge.
An automaton can traverse it in the direction $a$ without changing its state.
However, traversing it in the direction $-a$ requires at least $2(|D|-3)$ states,
where $D$ is the set of directions in the diode's signature.

If an automaton never moves in the direction $-a$,
then it can be transformed to an automaton with the same number of states,
operating on graphs in which every $(a, -a)$-edge is replaced with a diode.

Lower bound proofs for automata
with $n$ states over a signature with $k$ directions
use a diode designed for these particular values of $n$ and $k$.
This diode is denoted by $\Delta_{n,k}$.

For $n \geqslant 2$ and $k \geqslant 4$,
let $M = (4nk)!$,
and let $r=\lfloor\frac{k-2}{2}\rfloor$.
A diode $\Delta_{n,k}$ is defined over a signature $S_k$ that does not depend on $n$.

\begin{definition}
A signature $S_k = (D, -, \Sigma, \Sigma_0, (D_a)_{a \in \Sigma})$
consists of:
\begin{itemize}
\item
	the set of directions $D = \{a,-a\} \cup \{b_1, b_{-1}, \ldots, b_r, b_{-r}\}$;
\item
	opposite directions $-(a) = (-a)$, $-b_i = b_{-i}$, for $1 \leqslant i \leqslant r$;
\item
	the set of node labels
	$\Sigma = \{\mi{1}, \ldots, \mi{r}\} \cup \{\mi{-1}, \ldots, \mi{-r}\} \cup \{\m,\me,\ma\}$,
	with no initial labels defined ($\Sigma_0 = \emptyset$)
	since the diode is inserted into graphs;
\item
	sets of directions allowed at labels:
	$D_{\m} = D$,
	$D_{\mi{i}} = D_{\mi{-i}} = \{-a,b_i,-b_i\}$, for $i = 1, \ldots, r$,
	$D_{\me} = \{b_1,a,-a\}$,
	$D_{\ma} = \{-b_1,a\}$.
\end{itemize}
\end{definition}

A diode is comprised of $2r$ elements
$E_i, E_{-i}$, for $i \in \{1, \ldots, r\}$.
Each element $E_i$ and $E_{-i}$ is a graph over the signature $S_k$,
with two external edges, one with label $a$, the other with $-a$.
By these edges, the elements are connected in a chain.

\begin{figure}[t]
	\centerline{\includegraphics[scale=.9]{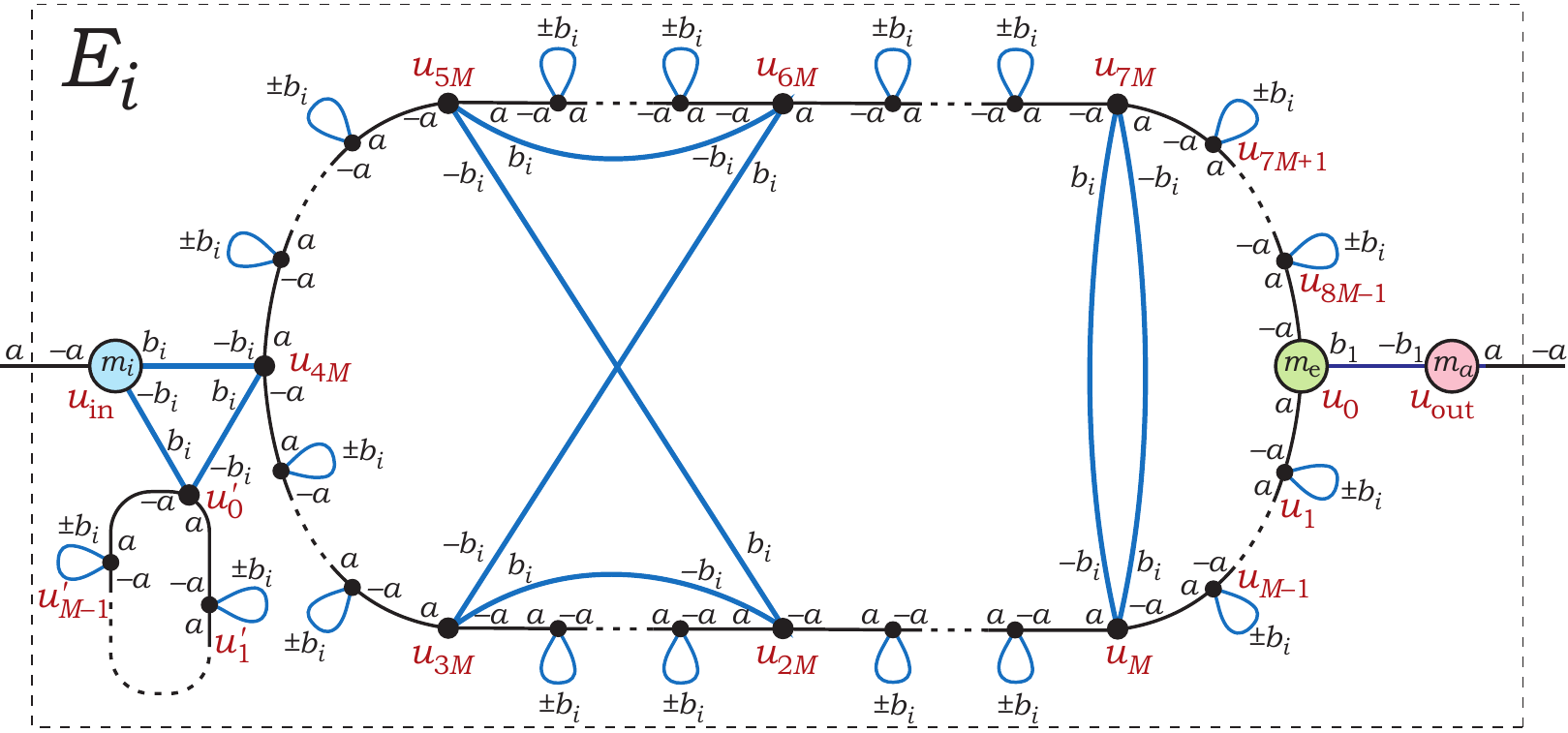}}
	\caption{Element $E_i$. Filled circles are nodes labelled with $\m$,
		each with $r-1$ loops in directions $\pm b_s$, with $s \neq i$.}
	\label{f:diode_element}
\end{figure}

The form of an element $E_i$ is illustrated in Figure~\ref{f:diode_element}.
Its main part is a cycle of length $8M$ in directions $a,-a$;
these are nodes $u_0, \ldots, u_{8M-1}$,
where the arithmetic in the node numbers is modulo $8M$,
e.g., $u_{-1} = u_{8M-1}$.
The node numbers are incremented in direction $a$.
Besides the main cycle,
there are two extra nodes: the entry point $u_{in}$ and the exit $u_{out}$,
as well as a small circle of length $M$
in directions $a,-a$ with the nodes $u_0', \ldots, u_{M-1}'$.
All nodes are labelled with $\m$,
except three: $u_{in}$ with label $\mi{i}$ matching the index of the element,
$u_{out}$ with label $\ma$,
and $u_0$ has label $\me$.

An element $E_i$ has specially defined edges in directions $b_i$ and $-b_i$.
Each node $u_j$ with $j \not\equiv 0 \pmod{M}$
has a $(b_i,-b_i)$-loop.
The nodes $u_j$ with $j \in \{M, 2M, 3M, 5M, 6M, 7M\}$
are interconnected with edges, as shown in Figure~\ref{f:diode_element};
these edges serve as traps for an automaton traversing the element backwards.
The node $u_{4M}$ has a different kind of trap
in the form of a cycle $u'_0$, \ldots, $u'_{M-1}$.
For all $s \neq i$, each node labelled with $\m$
has a $(b_s, -b_s)$-loop.

The element $E_{-i}$ is the same as $E_i$,
with the directions $b_i$ and $-b_i$ swapped.

\begin{figure}[t]
	\centerline{\includegraphics[scale=.9]{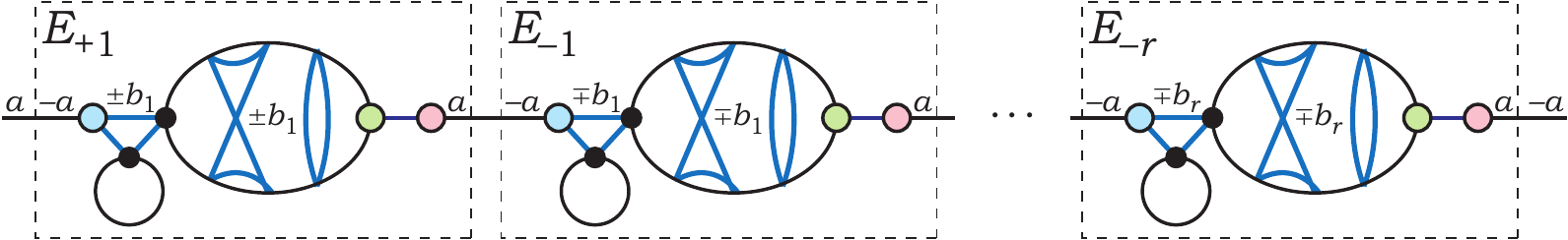}}
	\caption{Diode $\Delta_{n,k}$: a chain of elements $E_1$, $E_{-1}$, $E_2$, $E_{-2}$, \ldots, $E_r$, $E_{-r}$.}
	\label{f:diode}
\end{figure}

\begin{definition}
A diode element $E_i$, with $i \in \{\pm 1, \ldots, \pm r\}$,
is formally defined as follows.
\begin{itemize}
\item
The set of nodes is
$V = \{u_0, \ldots, u_{8M-1}\} \cup \{u_0', \ldots, u_{M-1}'\} \cup \{u_{in}, u_{out}\}$.

Node numbers are defined modulo $8M$.
For instance, $u_{-1}$ is a valid notation for $u_{8M-1}$.

\item
No initial node is defined, since a diode is a subgraph substituted into graphs,
and not a valid graph on its own.

\item
The nodes have the following labels.
\begin{align*}
	\lambda(u_{in}) &= \mi{i}
		\\
	\lambda(u_{out}) &= \ma
		\\
	\lambda(u_0) &= \me
\end{align*}
The rest of the nodes have label $\m$.

\item
The edges are defined as follows.
\begin{align*}
	u_{in} + (-a) &=
		\text{outside (point of entrance by $a$)}
		\\
	u_{in} + b_i &= u_{4M}
		\\
	u_{4M} + b_i &= u_0'
		\\
	u_0' + b_i &= u_{in}
		\\
	u_j' + a &= u_{j+1}'
		\\
	u_j' - a &= u_{j-1}'
		\\
	u_j' \pm b_s &= u_j',
		&& 
		\text{where }
		s \in \{1, \ldots, r\}, (u_j',b_s) \neq (u_0',\pm b_i)
		\\
	u_{out} + a &=
		\text{outside (point of exit by $a$)}
		\\
	u_{out} - b_1 &= u_0
		\\
	u_0+b_1 &= u_{out}
		\\
	u_j + a &= u_{j+1}
		\\
	u_j - a &= u_{j-1}
		\\
	u_j \pm b_s &= u_j,
		&& \text{where }
		s \in \{1, \ldots, r\} \setminus \{i, -i\}, j \neq 0
		\\
	u_j \pm b_i &= u_j,
		&& \text{where }
		j \notin \{0, M, 2M, 3M, 4M, 5M, 6M, 7M\}
		\\
	u_M \pm b_i &= u_{-M}
		\\
	u_{2M} + b_i &= u_{3M}
		\\
	u_{3M} + b_i &= u_{-2M}
		\\
	u_{-2M} + b_i &= u_{-3M}
		\\
	u_{-3M} + b_i &= u_{2M}
\end{align*}
\end{itemize}
\end{definition}

The diode $\Delta_{n,k}$
is a chain of such elements.

\begin{definition}
The diode $\Delta_{n,k}$, defined over a signature $S_k$,
is a chain of elements $E_i$,
joined sequentially as shown in Figure~\ref{f:diode}.
The order of elements is: $E_1, E_{-1}, E_2, E_{-2}, \ldots, E_r, E_{-r}$. 
For every element in the chain, there is an $a$-edge from its node $u_{out}$
to the node $u_{in}$ in the next element.
The entry point to the diode by $a$ is by the $-a$-edge of the first element $E_1$,
and the exit point is the $a$-edge of the last element $E_{-r}$.
\end{definition}

Each element can be traversed from the entrance to the exit
without changing the state:
at first, the automaton sees the label $m_i$, and accordingly moves in the direction $b_i$;
then, on labels $\m$, it proceeds in the direction $a$
until it reaches $u_0$, labelled with $\me$.
Then the automaton leaves the element by following directions $b_1$ and $a$.

The diode is hard to traverse backwards,
because the node $u_{4M}$ is not specifically labelled,
and in order to locate it,
the automaton needs to move in directions $\pm b_i$
from many nodes,
and is accordingly prone to falling into traps.

The diode is used as a subgraph
connecting two nodes of a graph as if an $(a, -a)$-edge.

\begin{definition}
For a graph $G$ over some signature $\widetilde{S}$,
let $G'$ be a graph obtained by replacing every $(a, -a)$-edge in $G$
with the diode $\Delta_{n,k}$.
Denote this graph operation by $h_{n,k} \colon G \mapsto G'$.
\end{definition}

The following lemma states that if an automaton never traverses an $(a, -a)$-edge backwards,
then its computations can be replicated on graphs with these edges substituted by diodes,
with no extra states needed.

\begin{lemma}\label{lemma_diod_forward}
Let $\widetilde{S}$ be any signature containing directions $a$, $-a$,
which has no node labels from the signature $S_k$.
Let $A = (Q,q_0,F,\delta)$ be a GWA
over the signature $\widetilde{S}$,
which never moves in the direction $-a$.

Then, there exists a GWA $A' = (Q,q_0,F,\delta')$
over a joint signature $\widetilde{S} \cup S_k$,
so that $A$ accepts a graph $G$ 
if and only if $A'$ accepts the graph $G'=h_{n,k}(G)$.
\end{lemma}
\begin{proof}
The automaton $A'$ uses the same states as $A$,
but the transition function is augmented with transitions by labels from the diode's signature.
Since none of the labels exists in the signature $\widetilde{S}$,
the new transitions would not contradict the existing ones.

For each state $q \in Q$, the following transitions are added to $\delta'$.
\begin{align*}
	\delta'(q,m_i) &= (q, b_i),
		&&\text{for }
		i \in \{\pm 1, \ldots, \pm r\}
		\\
	\delta'(q,\m) &= (q, a)
		\\
	\delta'(q,\me) &= (q, b_1)
		\\
	\delta'(q,m_a) &= (q, a)
\end{align*}

Let $G$ be any graph over the signature $\widetilde{S}$,
and let $G'=h_{n,k}(G)$ be the graph
constructed by replacing every $(a,-a)$-edge in $G$ with a diode.
Let $V$ be the set of nodes of $G$.
Then the set of nodes of $G'$, denoted by $V'$,
contains $V$ as a subset, and has extra nodes used by the substituted diodes.

Consider the computation of $A$ on the graph $G$.
It corresponds to the computation of $A'$ on $G'$ as follows.
Let $t_1, t_2, \ldots$ be the sequence of moments in the computation of $A'$ on $G'$,
at which the automaton visits any nodes from $V$.
It is claimed that each $i$-th step of the computation of $A$
corresponds to the step $t_i$ of the computation of $A'$,
as follows.

\begin{claim}
The automaton $A$, at the $i$-th step of its computation on $G$,
is in a node $v$ in a state $q$
	in and only if
$A'$, at the step $t_i$ on $G'$
is in the same node $v$ of $G'$, in the same state $q$.
\end{claim}

The claim is proved by induction on $i$.
The base case is step $i=0$,
when both $A$ and $A'$ are in their initial configurations.
Each configuration is of the form $(q_0, v_0)$.

Induction step.
Assume that both $A$ and $A'$ visit the same node $v$ 
in the same state $q$ at the moments $i$ and $t_i$, respectively.
The claim is that next, at the moments $i+1$ and $t_{i+1}$,
both automata are again in the same node and in the same state.
Let $f$ be the label of $v$.
There are the following cases.

\begin{itemize}
\item
	If this is an accepting pair $(q,f) \in F$,
	then, since $F = F'$, both automata accept.
\item
	If the transition $\delta(q,f)$ is undefined,
	then, since $v \in V$, its label is not from the diode's signature,
	and $A'$ has no new transitions at the label $\lambda(v)$.
	Then, the transition $\delta'(q, f)$ is undefined too,
	and both automata reject their graphs.
\item
	If the next transition is not in the directions $\pm a$,
	then $\delta(q,f) = \delta'(q,f) = (p,d)$,
	and both automata proceed to the next configuration $(p, v+d)$, where $v+d \in V$.
\item
	Let the next transition be in the direction $a$,
	so that $\delta(q,f) = (p,a)$. 
	At the next step, the automaton $A$ comes to the configuration $(p,u)$, 
	where $u$ denotes the node $v+a$ in the graph $G$.

For $A'$, the moment $t_{i+1}$ is the next visit to any of the nodes from $V$ after $t_i$.
At the next step after $t_i$, the automaton $A'$
enters the diode connecting the nodes $v$ and $u$ in the graph $G'$,
and changes its state to $p$, 
because $\delta'(q,f) = \delta(q,f) = (p,a)$. 
The transitions at the labels from the diode's signature
are defined so that the automaton $A'$ traverses the entire diode
in the state $p$, and leaves the diode for the node $u$. 
This $u$ is the next node from $V$ visited by $A'$ after $t_i$,
and $A'$ visits it in the state $p$, the same as the state of $A$.

\item
	The transition $\delta(q,f)$ cannot be in the direction $-a$,
	because, by the assumption, $A$ never moves in this direction.
\end{itemize}  

Thus, the correspondence between $i$ and $t_i$ has been established,
and then the automaton $A$ accepts the graph $G$
if and only if $A'$ accepts $G'$.
\end{proof}

Lemma~\ref{lemma_diod_forward}
shows that, under some conditions, a substitution of diodes
can be implemented on GWA without increasing the number of states.
The next lemma presents an \emph{inverse substitution of diodes}:
the set of pre-images under $h_{n,k}$ of graphs accepted by a GWA
can be recognized by another GWA with the same number of states.

\begin{lemma}\label{New_THE_LEMMA}
Let $k \geqslant 4$ and $n \geqslant 2$,
denote $h(G)=h_{n,k}(G)$ for brevity.
Let $\widetilde{S}$ be a signature
containing the directions $a,-a$
and no node labels from the diode's signature $S_k$.
Let $B = (Q,q_0,H,\sigma)$ be a GWA over the signature $\widetilde{S} \cup S_k$.
Then there exists an automaton $C = (Q,q_0,F,\delta)$ over the signature $\widetilde{S}$,
with the following properties.
\begin{itemize}
\item
	For every graph $G$ over $\widetilde{S}$,
	the automaton $C$ accepts $G$ if and only if $B$ accepts $h(G)$.
\item
	If $C$ can enter a state $q$ by a transition in direction $-a$,
	then $B$ can enter the state $q$ after traversing the diode backwards.
\item
	If $B$ is returning, then so is $C$.
\item
	If $B$ is halting, then $C$ is halting as well.
\end{itemize}
\end{lemma}

The automaton $C$ is constructed
by simulating $B$ on small graphs,
and using the outcomes of these computations
to define the transition function 
and the set of acceptance conditions of $C$.
Note that the signatures $\widetilde{S}$ and $S_k$
may contain any further common directions besides $a, -a$:
this does not cause any problems with the proof,
because the node labels are disjoint,
and thus $B$ always knows whether it is inside or outside a diode.

\begin{proof}[Proof of Lemma~\ref{New_THE_LEMMA}.]
The automaton $C = (Q,q_0,F,\delta)$ is constructed
by simulating $B = (Q,q_0,H,\sigma)$ on small graphs
and using the outcomes of these computations
to define the transition function $\delta$
and the set of acceptance conditions $F$.

\begin{figure}[t]
	\centerline{\includegraphics[scale=.9]{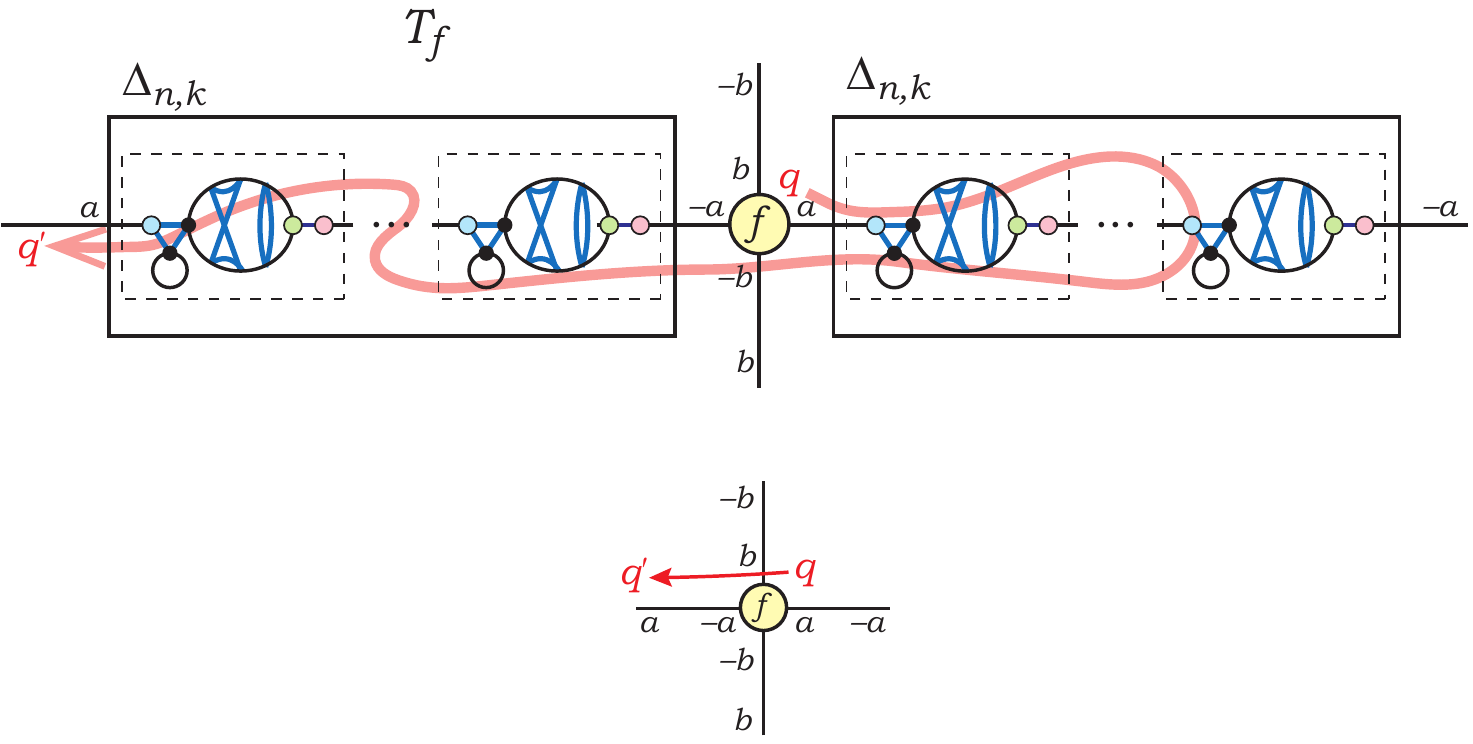}}
	\caption{The graph $T_f$ in the proof of Lemma~\ref{New_THE_LEMMA}.}
	\label{f:diode_T_f}
\end{figure}

For every state $q \in Q$ and label $f$ from $\widetilde{S}$,
the transition $\delta(q,f)$ is defined
by simulating the computation of $B$ on a graph denoted by $T_f$ and constructed as follows,
see Figure~\ref{f:diode_T_f}.
Let $v$ be a node with label $f$,
with attached edges in all directions in $D_f$, leading outside of $T_f$.
Edges in directions $a$ and $-a$ are replaced with diodes.
The automaton $B$ begins its computation at the node $v$ of $T_f$ in the state $q$.
If it eventually leaves $T_f$ by one of the edges leading outside,
in some direction $d$ and in a state $q'$,
then the transition is defined as $\delta(q,f) = (q',d)$.
If the automaton $B$ accepts without leaving $T_f$,
then the pair $(q,f)$ is accepting in $C$.
If $B$ loops or encounters an undefined transition,
then the transition $\delta(q,f)$ is undefined.

\begin{claim}
For every graph $G$ over the signature $\widetilde{S}$,
the automaton $C$ accepts $G$
if and only if
$B$ accepts $h(G)$.
Furthermore, if $B$ is returning, then $C$ is returning,
and if $B$ is halting, then $C$ is halting.
\end{claim}

\begin{proof}
Let $V$ be the set of nodes of $G$.
Then, $h(G)$ has the set of nodes $V \cup V_\Delta$,
where $V_\Delta$ is the set of all internal nodes in all diodes in $h(G)$.

Let $R_C = \set{(q_i,v_i)}{i = 0 ,\ldots, N_C}$ be the computation of $C$ on $G$, 
and let $R_B = \set{(p_j,u_j)}{j = 0 ,\ldots, N_B}$ be the computation of $B$ on $h(G)$.
The length of either computation, $N_C$ or $N_B$, can be infinite.

Each step of $C$ either repeats the corresponding step of $B$,
or contracts several moves of $B$ on $T_f$ into a single transition.
Let $m \colon \{0, \ldots, N_C\} \to \{0, \ldots, N_B\}$ be the function
that maps the number $i$ of a configuration in the computation $R_C$ 
to the number $j = m(i)$,
so that $(q_i,v_i) = (p_j,u_j)$ and $m(i+1)>m(i)$.
Note that configurations of two automata may be equal,
because $B$ and $C$ share the same set of states,
and every node of $G$ is a node of $h(G)$.

The function $m$ is constructed inductively on $i$.
\begin{itemize}
\item
	Basis: $i=0$.
	The configuration $(q_0,v_0)$ is initial in both computations,
	and $m(0) = 0$.
\item
	Induction step.
	Let $(q_i,v_i) = (p_{m(i)},u_{m(i)})$,
	and assume that it is not the last configuration in $R_C$.
	The goal is to find a number $m(i+1)$,
	such that $m(i) < m(i+1) \leqslant N_B$ and $(q_{i+1},v_{i+1}) = (p_{m(i+1)},u_{m(i+1)})$.

	Let $f$ be the label of $v_i$. 
	Then, $T_f$ is a subgraph of $h(G)$ centered around $v_i$.
	Since $\delta(q_i,v_i) = (q_{i+1}, d_{i+1})$, 
	where $d_{i+1}$ is the direction from $v_i$ to $v_{i+1}$ in the graph $G$, 
	the automaton $B$, having arrived to the node $v_i$ of $h(G)$ in the state $q_i$, 
	eventually leaves $T_f$ in the direction $d_{i+1}$ in the state $q_{i+1}$.
	The direction $d_{i+1}$ from $T_f$ leads to the node $v_{i+1}$.
	Thus, some time after the configuration $(q_i,v_i)$,
	the automaton $B$ reaches the configuration $(q_{i+1},v_{i+1})$.
	Let $m(i+1)$ be this moment in the computation $R_B$.
\end{itemize}

Now it will be shown that if $C$ accepts $G$, then $B$ accepts $h(G)$, 
and if $C$ does not accept $G$, then $B$ does not accept $h(G)$.

Let $C$ accept the graph $G$.
Consider the last configuration $(q,v)$
in the computation $R_C$.
It is accepting for $C$.
The automaton $B$ reaches the same configuration in its computation on $h(G)$.
Let $f$ be the label of $v$.
Consider the subgraph $T_{f}$ of $h(G)$ around the node $v$.
Since $(q,v)$ is an accepting configuration of $C$,
the automaton $B$ must accept $h(G)$ starting from the configuration $(q,v)$,
without leaving $T_f$.
So, $B$ accepts $h(G)$.
Note that if $B$ is returning,
then it can accept only in the initial node,
that is, $f = a_0$,
and then $C$ can accept only in the initial node as well.

Assume that $C$ does not accept $G$.
If $C$ loops, then $N_C = \infty$,
and in this case, since $m$ is an injection,
then $N_B = \infty$, and $B$ loops as well.
This, in particular, implies that if $B$ is halting, then $C$ is halting too.

The other case is when $C$ reaches a configuration $(q, v)$,
the label of $v$ is $f$, and $\delta(q, f)$ is undefined,
while $(q, f)$ is not an accepting pair.
Then the automaton $B$,
having started on $T_f$ at the node $v$ in the state $q$,
reaches an undefined transition without leaving $T_f$,
or loops inside $T_f$.
In both cases $B$ does not accept $h(G)$.
This confirms that if $C$ does not accept $G$, then $B$ does not accept $h(G)$.
\end{proof}

\begin{claim}
If $C$ can come to a state $q$ by a transition in the direction $-a$, 
then $B$ can come to $q$ after traversing a diode backwards.
\end{claim}
\begin{proof}
Let this transition in $C$ be
$\delta(p,f) = (q,-a)$.
By the construction of $C$,
this means that if $B$ runs on $T_f$ 
beginning in $v$ in the state $p$, 
then it leaves $T_f$ in the direction $-a$ in the state $q$.
Then, this transition is made after traversing the diode backwards.
\end{proof}
\end{proof}

The next lemma formally establishes that it is hard to traverse a diode backwards.

\begin{lemma}\label{lemma_diode_backward}
Let $A = (Q,q_0,F,\delta)$ be a GWA over a signature
that includes diode's signature $S_k$,
with $|Q| \leqslant 4nk$.
Assume that $A$, after traversing the diode $\Delta_{n,k}$ backwards,
can leave the diode in any of $h$ distinct states.
Then $A$ has at least $2h(k-3)$ states.
\end{lemma}

While moving through an element $E_i$ backwards,
the automaton sees labels $\m$ most of the time,
and soon begins repeating a periodic sequence of states.
Without loss of generality, assume that this periodic sequence
contains more transitions in the direction $a$ than in $-a$.
Then the automaton reaches the node $u_M$,
and at this point it may teleport between $u_M$ and $u_{-M}$ several times.
Let $w \in \{b_i, -b_i\}^*$ be the sequence of these teleportation moves,
and let $x$ be the corresponding sequence of states. 
Depending on the sequence $w$,
the automaton may eventually exit the cycle to the node $u_{in}$,
or fall into one of the traps and get back to $u_0$.
It is proved that for the automaton to reach $u_{in}$,
the string $w$ must be non-empty and of even length;
furthermore, if $|w|=2$, then $w=(-b_i)b_i$.

Now consider the $h$ backward traversals of the diode ending in some states $p_1$, \ldots, $p_h$.
When the traversal ending in $p_j$ proceeds through the element $E_i$,
the strings $w_{i,j} \in \{b_i, -b_i\}^*$ and $x_{i,j}$ are defined as above.
Then, as the last step of the argument, it is proved that whenever $|w_{i,j}|=2$,
the states in $x_{i,j}$ cannot occur in any other string $x_{i',j'}$.
For $w_{i,j}$ of length 4 or more,
the states in $x_{i,j}$ can repeat in other strings $x_{i',j'}$, but only once.
It follows that there are at least $2h(k-3)$ distinct states in these strings.

\begin{proof}[Proof of Lemma~\ref{lemma_diode_backward}.]
Assume that $A$ moves through the element $E_i$
from the node $u_{out}$ to the node $u_{in}$.
In this computation, there is the moment when $A$ visits $u_0$ for the last time
before leaving the element $E_i$.
Let $t_0$ denote this moment, as the number of a computation step.
At this moment, the automaton makes a transition
that puts it either on the segment from $u_0$ to $u_M$,
or on the segment from $u_0$ to $u_{-M}$.
Since, by assumption, the automaton shall never return to $u_0$
and shall eventually reach $u_{in}$,
it must traverse this segment and reach the corresponding node $u_{\pm M}$.
Let $t_1$ be the moment of the first visit to any of the nodes $u_{\pm M}$
after $t_0$.

Since the moment $t_0$
and until the subsequent first visit to $u_{in}$,
the automaton shall move over nodes labelled with $\m$.
Therefore, already after $|Q|$ steps or earlier---%
and much before the moment $t_1$, since $M \gg |Q|$---%
the automaton loops, that is, begins repeating
some sequence of states $q_1, \ldots, q_p$, 
and some sequence of directions $d_1, \ldots, d_p$,
where $p$ is the minimal period,
and the enumeration of states in the period
is chosen so that at the moment $t_1$ the automaton is in the state $q_1$.

In the sequence $d_1, \ldots, d_p$,
one of the directions $a,-a$ must occur more often than the other,
since otherwise the automaton never reaches the node $u_{\pm M}$.
Assume that the direction $a$ is the one that occurs more often
(the case of $-a$ is symmetric),
and the automaton accordingly moves from $u_0$ to $u_M$.
On the way, the automaton actually moves only in directions $\pm a$,
while all its transitions in directions $\pm b_j$
follow the loops and do not move the automaton.
However, as the automaton moves through the node $u_M$,
the transitions in directions $\pm b_i$ are executed,
and the automaton, without noticing that,
teleports to the node with an opposite number.
While following the sequence of directions $d_1, \ldots, d_p$,
the automaton may move away from the nodes $u_{\pm M}$
by several edges in directions $a$ and $-a$,
then get back, get teleported to the other part of the graph in the directions $\pm b_i$,
then again move away, etc.
But since $a$ occurs in the periodic part more often than $-a$,
eventually the automaton passes through the pair of nodes $u_{\pm M}$
and moves on.

Let $w \in \{b_i, -b_i\}^*$ be the string of directions,
in which the automaton,
having arrived to the node $u_M$ in the state $q_1$,
teleports between the nodes $u_M$ and $u_{-M}$,
until the general flow of its motion in the direction $a$ 
moves it away from this pair of nodes.
Let $x$ be the corresponding string of states,
in which the automaton makes the transitions in the string $w$.

If the automaton teleports between the nodes $u_M$ and $u_{-M}$ an even number of times,
that is, if $|w|$ is even,
then it proceeds further to the segment from $u_M$ to $u_{2M}$.
If the length of $w$ is odd, then the automaton is teleported to the node $u_{-M}$
and sets foot on the path in the direction of $u_0$:
and then, moving in the general direction $a$,
it reaches the node $u_0$, which contradicts the assumption.
Thus, it has been proved that the number of teleportations is even.

\begin{claim}\label{lemma_diode_backward__w_even}
The length of $w$ is even,
and the automaton, having arrived to the node $u_M$ in the state $q_1$,
continues in the direction of the node $u_{2M}$.
\end{claim}

Let $s$ be the difference between the number of occurrences
of $a$ and $-a$ in the sequence $d_1, \ldots, d_p$.
Since $s \leqslant p \leqslant |Q|$, the number $M$ is divisible by $s$,
and thus, from the moment $t_1$ of the first visit to $u_M$
and until the moment of the first visit to $u_{2M}$,
the automaton makes a whole number of periods $\frac{M}{s}$,
and accordingly comes to $u_{2M}$ in the state $q_1$.

If the string $w$ is empty,
then the automaton moves directly through $u_{2M}$,
without teleporting anywhere in the directions $\pm b_i$,
and then passes by $u_{iM}$, for $i=3,4,5,6,7$, in the same way,
getting to each of these nodes in the state $q_1$.
Eventually, contrary to the assumption, it comes to the node $u_0$.
Therefore, this case is impossible.

\begin{claim}\label{lemma_diode_backward__w_nonempty}
The string $w$ is non-empty.
\end{claim}

Thus, having reached the node $u_{2M}$,
the automaton, following directions from a non-empty string $w$ of even length,
teleports several times between the nodes $u_{2M}$, $u_{3M}$, $u_{5M}$ and $u_{6M}$.
In the following, it will be proved that if $w$ is of length 2,
then it is uniquely defined.

\begin{claim}\label{lemma_diode_backward__w_length_2}
If $|w|=2$, then $w = (-b_i)b_i$.
\end{claim}

Indeed, if $w = b_i b_i$ or $w = (-b_i) (-b_i)$,
then the automaton, while passing through the node $u_{2M}$,
teleports to the node $u_{6M}$,
and then, gradually moving in the general direction $a$,
reaches the node $u_0$. This is a contradiction.

If $w = b_i (-b_i)$, then the automaton
passes through the nodes $u_{iM}$ for $i=3,4,5,6,7$,
first getting into each of these nodes in the state $q_1$,
and finally comes to the node $u_0$.
It should be mentioned
that, while passing through the node $u_{4M}$,
the automaton teleports to the node $u'_0$,
but since its vicinity in the directions $\pm a$
is indistinguishable from long segments of the large cycle,
the automaton does not see any difference
and eventually teleports back to $u_{4M}$,
without ever paying a visit in the direction $b_i$
and without seeing the node $u_{in}$.

Therefore, there is only one possibility left, that $w=(-b_i) b_i$,
or otherwise $w$ must be of length at least 4.

In the case when $-a$ occurs in the period more often than $a$,
Claims~\ref{lemma_diode_backward__w_even}--\ref{lemma_diode_backward__w_length_2}
on the properties of the string $w$ hold as stated and can be proved analogously.

By the assumption, there are $h$ backward traversals of the diode
ending in $h$ distinct states: $p_1$, \ldots, $p_h$.
Each of these traversals includes passing through each element $E_i$,
for $i \in \{\pm 1, \ldots, \pm r\}$.
For the computation that eventually leaves the diode in the state $p_j$,
one can define the string $w_{i,j} \in \{b_i, -b_i\}^*$ of directions
in which the automaton ``teleports'',
as $w$ was defined above.
Let $x_{i,j}$ be the corresponding string of states.
All properties of these strings hold as stated.

It will be proved that in the sequences of states $x_{i,j}$,
for $i = \pm 1, \ldots, \pm r$ and $j = 1, \ldots, h$,
there are in total at least $4rh$ distinct states,
which will prove the lemma.

First, consider that in the states from $x_{i,j}$,
the automaton moves in the directions listed in $w_{i,j}$,
while $w_{i,j}$ contains only directions $b_i,-b_i$.
Then, as long as $i_1 \neq i_2$ and $i_1 \neq -i_2$,
the sets of states used in $x_{i_1,j}$ and $x_{i_2,j}$ are disjoint.
Then, it is sufficient to prove that, for each $i$,
the strings $x_{i,j},x_{-i,j}$, for all $j = 1, \ldots, h$,
together contain at least $4h$ distinct states.

It is claimed that, for each $i \in \{\pm 1, \ldots, \pm r\}$,
all states in the strings $x_{i,1}, \ldots, x_{i,h}$ are pairwise distinct.
Indeed, suppose that some state $q$ occurs twice in these strings.
Within a single string $x_{i,j}$, all states are known to be distinct.
Then, there exist different $j$ and $j'$,
such that $q$ occurs both in $x_{i,j}$ and in $x_{i,j'}$.
By definition, if the state $q$ is in the string $x_{i,j}$,
this means that the computation through $E_i$
passes through one of the nodes $u_{M}$ and $u_{-M}$ in the state $q$.
This computation does not return to $u_0$ anymore,
and eventually leads the automaton
out of the diode in the state $p_j$;
and if the automaton is put in the same state $q$
into the other of the two nodes $u_M$ and $u_{-M}$,
then it returns to the node $u_0$ of the element $E_i$.
Since $q$ also occurs in the string $x_{i,j'}$,
then, by the same reasoning,
the computation starting in one of the nodes $u_{M}$ and $u_{-M}$
proceeds out of the diode in the state $p_{j'}$.
This is a contradiction, which establishes the following claim.

\begin{claim}
All states in the strings $x_{i,1}, \ldots, x_{i,h}$ are distinct,
and all states in the strings $x_{-i,1}, \ldots, x_{-i,h}$ are distinct as well.
\end{claim}

Note that the strings $x_{i,j}$ and $w_{i,j}$
can be uniquely reconstructed from each state $q$ from $x_{i,j}$.
Indeed, a state $q$ is a part of a periodic sequence of states on the labels $\m$.
Then the automaton $A$ can be put in the state $q$ into one of the nodes $u_M$, $u_{-M}$---%
the one from which it will not return to the node $u_0$ of $E_i$.
Then, the cycle can be unrolled forward and backward,
and the periodic part of the computation is thus reconstructed.
And then, the strings of states and directions $x_{i,j}$ and $w_{i,j}$
are extracted from this periodic part.

The strings $w_{i,j}$ can be of two kinds:
either $w_{i,j} = (-b_i)b_i$, or $|w_{i,j}| \geqslant 4$.
If a string $w_{i,j}$ is of length 2,
then it cannot coincide with any of the strings $w_{-i,j'}$,
because $w_{i,j} = (-b_i)b_i$,
whereas $w_{-i,j'}$ is either equal to $b_i(-b_i)$, or is of length at least 4.
Since the strings $x$ and $w$ are reconstructed from a single state,
if a string $x_{i,j}$ is of length 2,
then its states do not occur in any other strings $x_{i',j'}$.

Now it can be proved, for each $i$, that there are at least $4h$ distinct states
in the strings $x_{i,j}$, $x_{-i,j}$, for $j = 1, \ldots, h$.
All states in the strings $x_{i,1}, \ldots, x_{i,h}$ are pairwise distinct;
so are the states in the strings $x_{-i,1}, \ldots, x_{-i,h}$.
Therefore, every state can occur at most twice:
once in one of the strings $x_{i,1}, \ldots, x_{i,h}$,
and the other time in one of the strings $x_{-i,1}, \ldots, x_{-i,h}$.
Let $c$ be the number of strings of length 2
among the $2h$ strings $x_{\pm i,j}$, with $j = 1, \ldots, h$.
Then, there are $2h-c$ strings of length at least 4.
All states occurring in the strings of length 2 are unique;
states in the rest of the strings can coincide, but only in pairs.
Overall, there are no fewer than $2c+\frac{4(2h-c)}{2} = 4h$ distinct states,
as desired.
\end{proof}

\section{Lower bound on the size of returning automata}\label{section_returning}

By the construction of Kunc and Okhotin~\cite{KuncOkhotin_reversible},
as improved in Section~\ref{section_upper_bounds},
an $n$-state GWA over a signature with $k$ directions
can be transformed to a returning GWA with $2nk+n$ states.
A closely matching lower bound will now be proved
by constructing an automaton with $n$ states
over a signature with $k$ directions,
such that every returning automaton
that recognizes the same set of graphs
must have at least $2(n-1)(k-3)$ states.

The first step is a construction of a simple automaton
over a signature $\widetilde{S}$ with four directions $a, -a, b, -b$
and two graphs over this signature,
so that the automaton accepts one of them and rejects the other.
The automaton will have $n$ states,
it will never move in the direction $-a$,
and every returning automaton recognizing the same set of graphs
can enter $n-1$ distinct states after transitions in the direction $-a$.
Then, Lemma~\ref{lemma_diode_backward} shall assert
that every returning automaton
recognizing the same graphs with diodes substituted
must have the claimed number of states.

\begin{definition}
The signature $\widetilde{S} = (D, -, \Sigma, \Sigma_0, (D_a)_{a \in \Sigma})$
uses
	the set of directions $D = \{a,-a,b,-b\}$,
	with $-(a) = (-a)$, $-(b) = (-b)$.
	The set of node labels is $\Sigma = \{c_0, c,c_l,c_r,c_{acc}\}$,
	with initial labels $\Sigma_0 = \{c_0\}$.
	The allowed directions are 
$D_{c} = D$,
$D_{c_0} = D_{c_l} = \{a\}$, 
and
$D_{c_r} = D_{c_{acc}} = \{-a\}$.
\end{definition}

\begin{figure}[t]
	\centerline{\includegraphics[scale=.9]{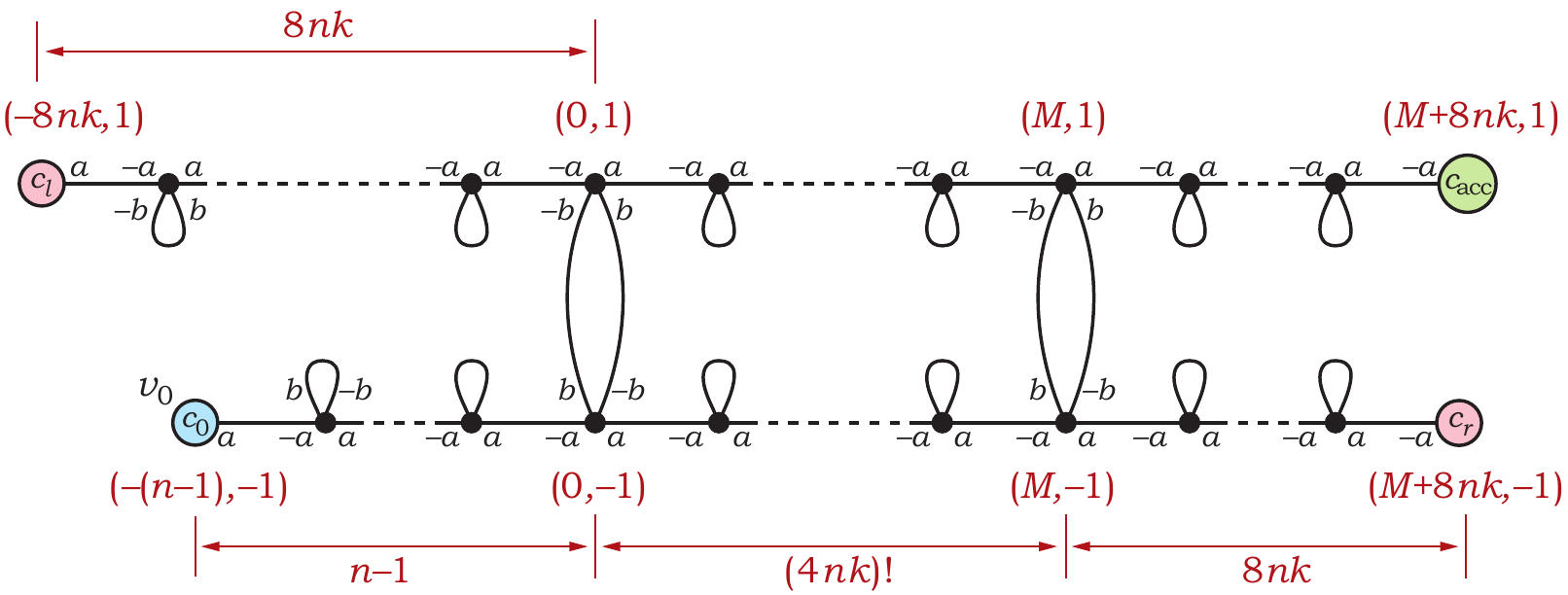}}
	\caption{The graph $G_{n,k}^{accept}$.}
	\label{f:gwa_returning_and_reversible_lower_bound_witness}
\end{figure}

For $n \geqslant 2$ and $k \geqslant 4$,
let $M=(4nk)!$ be as in the definition of the diode $\Delta_{n,k}$.
Let $G_{n,k}^{accept}$ and $G_{n,k}^{reject}$
be two graphs over the signature $\widetilde{S}$,
defined as follows.
The graph $G_{n,k}^{accept}$
is illustrated in Figure~\ref{f:gwa_returning_and_reversible_lower_bound_witness};
the other graph $G_{n,k}^{reject}$ is almost identical,
but the node that determines acceptance is differently labelled.

Both graphs consist of two horizontal chains of nodes, connected by bridges at two places.
Nodes are pairs $(x,y)$,
where $y \in \{-1, 1\}$ is the number of the chain,
and $x$ is the horizontal coordinate,
with $-(n-1) \leqslant x \leqslant M+8nk$ for the lower chain ($y=-1$)
and $-8nk \leqslant x \leqslant M+8nk$ for the upper chain ($y=1$).

All nodes except the ends of chains have labels $c$.
The node $(-(n-1),-1)$ is the initial node, with label $c_0$.
The other left end $(-8nk,1)$ is labelled with $c_l$. 
The node $(M+8nk,-1)$ has label $c_r$.
The node $(M+8nk,1)$
is labelled with $c_{acc}$ in $G_{n,k}^{accept}$
and with $c_r$ in $G_{n,k}^{reject}$;
this is the only difference between the two graphs.

The horizontal chains are formed of $(a,-a)$-edges,
with $a$ incrementing $x$ and $-a$ decrementing it.
Edges with labels $(b,-b)$ are loops at all nodes
except for $(0,1)$, $(0,-1)$, $(M,1)$ and $(M,-1)$.
The latter four nodes form two pairs connected with bridges in directions $(b,-b)$.

\begin{definition}
The graphs $G_{n,k}^{accept}$ and $G_{n,k}^{reject}$
over the signature $\widetilde{S}$ are defined as follows.

\begin{itemize}
\item
The set of nodes is
$V = \setbig{(x,-1)}{x \in \{-(n-1), \ldots, M+8nk\}} \cup \setbig{(x,1)}{x \in \{-8nk, \ldots, M+8nk\}}$.

\item
	The initial node is $v_0 = (-(n-1),-1)$.

\item
The labels of the nodes are as follows.
\begin{align*}
	\lambda((x,y)) &= c,
		\text{except for }
		(x,y) \in \{(-(n-1),-1), (M+8nk,-1), (-8nk,1), (M+8nk,1)\}.
		\\
	\lambda((-(n-1),-1)) &= c_0
		\\
	\lambda((M+8nk,-1)) &= c_r
		\\
	\lambda((-8nk,1)) &= c_l
		\\
	\lambda((M+8nk,1)) &= \begin{cases}
		c_{acc},
			& 
			\text{for }
			G_{n,k}^{accept}
				\\
		c_r,
			& 
			\text{for }
			G_{n,k}^{reject}
	\end{cases}
\end{align*}

\item
	The following edges are defined.
\begin{align*}
	(x,y) + a &= (x+1,y),
		&&
		\text{if }
		(x,y) \neq (M+8nk,-1), (x,y) \neq (M+8nk,1) 
		\\
	(x,y) + (-a) &= (x-1,y),
		&&
		\text{if }
		(x,y) \neq (-(n-1),-1), (x,y) \neq (-8nk,1)
		\\
	(x,y) + b &= (x,y),
		&&
		\text{if }
		(x,y) \notin \{(M+8nk,\pm 1), (-8nk,1), (-(n-1),-1)\}, x \neq 0, x \neq M
		\\
	(x,y) + (-b) &= (x,y),
		&&
		\text{if }
		(x,y) \notin \{(M+8nk,\pm 1), (-8nk,1), (-(n-1),-1)\}, x \neq 0, x \neq M
		\\
	(0,y) + b &= (0,-y)
		\\
	(0,y) +(- b) &= (0,-y)
		\\
	(M,y) + b &= (M,-y)
		\\
	(M,y) + (-b) &= (M,-y)
\end{align*}
\end{itemize}
\end{definition}

An $n$-state automaton $A$, that accepts the graph $G_{n,k}^{accept}$,
does not accept any graphs without labels $c_{acc}$
and never moves in the direction $-a$,
is defined as follows.
In the beginning, it moves in the direction $a$ in the same state $q_0$,
then makes $n-2$ further steps in the direction $a$, incrementing the number of state.
Next, it crosses the bridge in the direction $b$
and enters the last, $n$-th state,
in which it moves in the direction $a$ until it sees the label $c_{acc}$.

\begin{definition}\label{automaton_A_definition}
The graph-walking automaton $A = (Q, q_0, \delta, F)$ over the signature $\widetilde{S}$
consists of:
\begin{itemize}
\item
	the set of states
	$Q = \{q_0,\ldots,q_{n-1}\}$;
\item
	the initial state $q_0$;
\item
	the transition function $\delta$, defined by
	\begin{align*}
	\delta(q_0,c_0) &= (q_0,a) 
		\\
	\delta(q_i,c) &= (q_{i+1},a),
		&& \text{if }
		i \in \{0, \ldots, n-3\}
		\\	
	\delta(q_{n-2},c) &= (q_{n-1},b)
		\\	
	\delta(q_{n-1},c) &= (q_{n-1},a)
	\end{align*}
\item
	the set of acceptance conditions
	$F = \{(q_{n-1},c_{acc})\}$.
\end{itemize}
\end{definition}

\begin{lemma}\label{returning_automaton_reaches_n_minus_1_states_by_minus_a_lemma}
Every returning automaton $A'$
that accepts the same set of graphs as $A$, 
and has at most $4nk$ states,
can enter at least $n-1$ distinct states
after transitions in the direction $-a$.
\end{lemma}

The proof of Lemma~\ref{returning_automaton_reaches_n_minus_1_states_by_minus_a_lemma}
is inferred from the following lemma.

\begin{lemma}\label{backwards_traversal_requires_n_minus_1_by_a_and_n_minus_1_by_minus_a_lemma}
Let $n \geqslant 2$ and $k \geqslant 4$.
Let an automaton with at most $4nk$ states
operate on a graph $G_{n,k}^{accept}$.
Assume that it begins its computation
at one of the ends of the upper chain, $(-8nk, 1)$ or $(M+8nk, 1)$,
and assume that it arrives at
one of the ends of the lower chain, $(-(n-1), -1)=v_0$ or $(M+8nk, -1)$,
without visiting either end of the upper chain on the way.
Then the automaton must arrive to $v_0$,
and the periodic sequence of directions
which it ultimately follows
contains at least $n-1$ moves in the direction $a$
and at least $n-1$ moves in the direction $-a$.
\end{lemma}
\begin{proof}
All nodes visited by the automaton during this computation
are labelled with $c$,
and early in the computation it begins repeating the same periodic sequence of actions.
Let $d_1 \ldots d_p$, with $p \leqslant 4nk$,
be the periodic sequence of directions.

The proof is slightly different for computations
beginning at the node $(-8nk, 1)$ and at the node $(M+8nk, 1)$.

\subparagraph{Case 1: the automaton begins at the node $(M+8nk,1)$.}
Since the automaton eventually arrives at a coordinate $x \leqslant M$,
the sequence $d_1 \ldots d_p$ has more occurences of $-a$ than of $a$.
Hence, it is sufficient to show
that the number of occurences of $a$ is at least $n-1$.
The proof is by contradiction.
Suppose that there are at most $n-2$ occurrences of $a$.
Let $s$ be the difference between the number of occurrences of $-a$ and $a$.
It is claimed that the automaton visits the node $(-8nk,1)$,
which would contradict the assumption
that it never returns to either end of the upper chain.

Moving periodically, the automaton moves in directions $a$ and $-a$,
shifting by $s$ edges to the left at each period,
while also applying transitions in directions $b$ and $-b$,
which at first follows the loops.
When the automaton reaches the point $x=M$ along the $x$ axis,
the same transitions may change its $y$-coordinate.
As the automaton continues shifting horizontally,
it will eventually reach $(M-(n-1),1)$ in the upper chain
or $(M-(n-1),-1)$ in the lower chain.
It is claimed that, by this moment,
it will have moved far enough from $x=M$,
so that it would not return to that position on its way from right to left,
and continue moving along the current chain.
Indeed, since, by the assumption, the sequence $d_1 \ldots d_p$
has at most $n-2$ moves in the direction $a$,
there is no way the automaton could get back.

Assume that the automaton remains on the upper chain,
that is, passes through the node $(M-(n-1),1)$,
and does not visit $(M-(n-1),-1)$.
Since the number $M=(4nk)!$ is divisible by $s$,
the automaton reaches the coordinate $x=0$
in the same state in which it has earlier arrived to the coordinate $x=M$.
This means that it repeats the same transitions as before
and again stays on the upper chain,
that is, comes to the node $(-(n-1), 1)$,
without paying a visit to the node $(-(n-1), -1)=v_0$.
Therefore, it continues its periodic motion until it comes to the node $(-8nk,1)$.
A contradiction has been obtained.

The other possibility is that the automaton passes through the node $(M-(n-1),-1)$
without visiting the node $(M-(n-1),1)$,
and thus moves to the lower chain.
The automaton arrives to the coordinate $x=0$
in the same state as it arrived to $x=M$.
It repeats the same transitions
and again moves to the other chain,
this time back to the upper chain.
Then, as in the previous case, it arrives to the node $(-(n-1), 1)$
without visiting $(-(n-1), -1)=v_0$.

\subparagraph{Case 2: the automaton begins at the node $(-8nk,1)$.}
The automaton soon begins repeating
a periodic sequence of directions $d_1 \ldots d_p$,
with $p \leqslant 4nk$.
This time, moves in the direction $a$ are more frequent than moves in the direction $-a$.
Let $s$ be their difference. 
This time it is enough to prove
that moves in the direction $-a$
occur in the sequence at least $n-1$ times.

The proof is by contradiction.
Suppose that the sequence has fewer than $n-1$ occurrences of $-a$.
Then, gradually shifting from left to right
the automaton cannot reach the node $(-(n-1), -1)=v_0$.

While passing through the node $(0, 1)$,
the automaton may continue on the upper chain
or move to the lower chain.
In both cases it arrives to a node with coordinate $x=M$
in the same state in which it came to the coordinate $x=0$.
If it moved to the lower chain at the first time,
it returns to the upper chain at the second time;
and if it stayed on the upper chain at the first time,
it stays on it at the second time.
Thus, the automaton continues on the upper chain 
and arrives to the node $(M+8nk, 1)$.
This contradicts the assumption
that the automaton never returns to any end of the upper chain.
\end{proof}

\begin{proof}[Proof of Lemma~\ref{returning_automaton_reaches_n_minus_1_states_by_minus_a_lemma}]
Consider the computation of $A'$ on the graph $G_{n,k}^{accept}$.
The automaton $A'$ must visit the node $(M+8nk,1)$, 
since its label is the only difference between the graphs $G_{n,k}^{accept}$ and $G_{n,k}^{reject}$.
Denote by $t_0$ the moment of the last visit
to any of the nodes $(M+8nk,1)$ and $(-8nk,1)$.
At some point after $t_0$,
the automaton starts behaving periodically,
and, by Lemma~\ref{backwards_traversal_requires_n_minus_1_by_a_and_n_minus_1_by_minus_a_lemma},
there are at least $n-1$ moves in the direction $-a$ in the periodic sequence of directions.
Since all states in the periodic part are pairwise distinct,
the automaton $A'$ enters at least $n-1$ distinct states
after transitions in the direction $-a$.
\end{proof}

It remains to combine this lemma
with the properties of the diode
to obtain the desired theorem.

\begin{theorem}\label{theorem_ret}
For every $k \geqslant 4$,
there exists a signature with $k$ directions,
such that, for every $n \geqslant 2$,
there is an $n$-state graph-walking automaton,
such that every returning automaton recognizing the same set of graphs
must have at least $2(n-1)(k-3)$ states.
\end{theorem}

\begin{proof}
The proof uses the automaton $A$ defined above.
By Lemma~\ref{lemma_diod_forward},
the $n$-state automaton $A$ over the signature $\widetilde{S}$,
is transformed to $n$-state automaton $A'$ over the signature $\widetilde{S} \cup S_k$.
The directions $\pm a$ are the same for $\widetilde{S}$ and $S_k$,
and $\pm b$ in $\widetilde{S}$
are merged with $\pm b_1$ in $S_k$,
so there are $k$ directions in total.

For every graph $G$,
the automaton $A'$ accepts a graph $h_{n,k}(G)$
with $(a,-a)$-edges replaced by diodes,
if and only if
$A$ accepts $G$.
The automaton $A'$ is the desired example:
it is claimed that every returning automaton $B$
recognizing the same set of graphs as $A'$
has at least $2(n-1)(k-3)$ states.

Let $B$ be any returning automaton with at most $4nk$ states recognizing these graphs.
By Lemma~\ref{New_THE_LEMMA},
there is an automaton $C$ over the signature $\widetilde{S}$
and with the same number of states,
which accepts a graph $G$ if and only if $B$ accepts $h(G)$.
This is equivalent to $A$ accepting $G$,
and so $C$ and $A$ accept the same set of graphs.
Since $B$ is returning,
by Lemma~\ref{New_THE_LEMMA}, $C$ is returning too.
Then, Lemma~\ref{returning_automaton_reaches_n_minus_1_states_by_minus_a_lemma} asserts that
the automaton $C$ may enter $n-1$ distinct states
after moving in the direction $-a$.

Then, according to Lemma~\ref{New_THE_LEMMA},
the automaton $B$ enters at least $n-1$ distinct states
after traversing the diode backwards.
Therefore, by Lemma~\ref{lemma_diode_backward},
this automaton should have at least $2(k-3)(n-1)$ states.
\end{proof}

\section{Lower bound on the size of halting automata}\label{section_halting}

Every $n$-state GWA with $k$ directions
can be transformed to a halting GWA with $2nk+1$ states,
as shown in Section~\ref{section_upper_bounds}.
In this section,
the following lower bound for this construction is established.

\begin{theorem}\label{theorem_stop}
For every $k \geqslant 4$, there is a signature with $k$ directions,
such that for every $n \geqslant 2$
there is an $n$-state GWA,
such that every halting automaton accepting the same set of graphs
has at least $2(n-1)(k-3)$ states.
\end{theorem}

The argument shares some ideas
with the earlier proof for the case of returning automata:
the signature $\widetilde{S}$,
the graphs $G_{n,k}^{accept}$ and $G_{n,k}^{reject}$,
and the automaton $A$
are the same as constructed in Section~\ref{section_returning}.
The proof of Theorem~\ref{theorem_stop} uses the following lemma,
stated similarly to Lemma~\ref{returning_automaton_reaches_n_minus_1_states_by_minus_a_lemma}
for returning automata.

\begin{lemma}\label{stopping_automaton_reaches_n_minus_1_states_by_minus_a_lemma}
Every halting automaton $A'$, 
accepting the same set of graphs as $A$
and using at most $4nk$ states,
must enter at least $n-1$ distinct states
after transitions in the direction $-a$.
\end{lemma}
\begin{proof}
Consider the computation of $A'$ on the graph $G$,
defined by modifying $G_{n,k}^{reject}$ as follows:
the nodes $(M+8nk,1)$ and $(-8nk,1)$
are merged into a single node $v_{joint}$, with label $c$,
and with a loop by $\pm b$.

\begin{figure}[t]
	\centerline{\includegraphics[scale=.9]{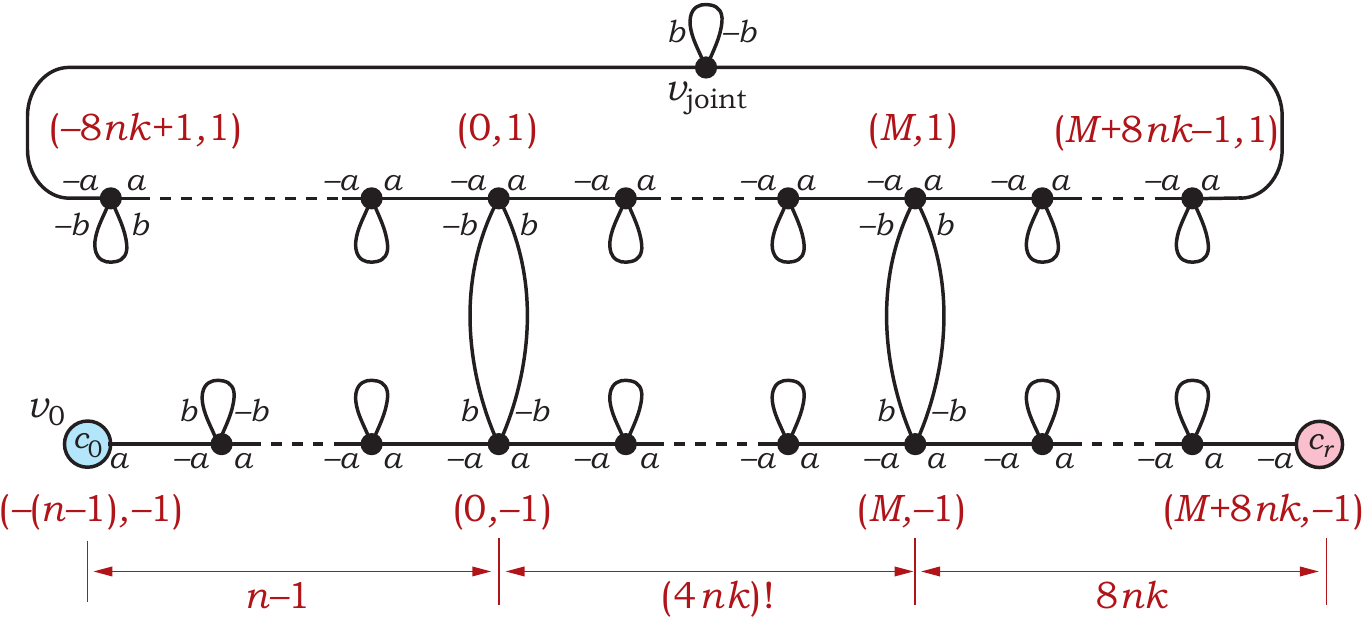}}
	\caption{The graph $G$ in the proof of Lemma~\ref{stopping_automaton_reaches_n_minus_1_states_by_minus_a_lemma}.}
	\label{f:gwa_returning_and_reversible_lower_bound_witness__joined}
\end{figure}

The automaton $A'$ must visit the node $v_{joint} = (M+8nk,1) = (-8nk,1)$, 
because this node is the only difference between $G$ and $G_{n,k}^{accept}$. 
By the time the automaton reaches this node,
it already executes a periodic sequence of states 
and directions. 
Since $A'$ should halt
at some time after visiting $v_{joint}$,
it needs to stop its periodic behaviour,
which requires visiting any label other than $c$.
Hence, the automaton should reach one of the ends of the lower chain.

The argument in the proof of Lemma~\ref{backwards_traversal_requires_n_minus_1_by_a_and_n_minus_1_by_minus_a_lemma}
is applicable to the segment of the computation
from the last visit of $v_{joint}$
until arriving to one the ends of the lower chain.
Then, the periodic sequence of states on this segment
should contain at least $n-1$ occurrences of states
reached after a transition in the direction $-a$.
\end{proof}

\begin{proof}[Proof of Theorem~\ref{theorem_stop}]
The proof is analogous to the proof of Theorem~\ref{theorem_ret}.
Lemma~\ref{stopping_automaton_reaches_n_minus_1_states_by_minus_a_lemma}
is used instead of Lemma~\ref{returning_automaton_reaches_n_minus_1_states_by_minus_a_lemma};
and the application of Lemma~\ref{New_THE_LEMMA}
now uses the preservation of the halting property. 
\end{proof}

\section{Lower bound on the size of returning and halting automata}\label{section_returning_and_halting}

An $n$-state GWA over a signature with $k$ directions
can be transformed to an automaton
that \emph{both} halts on every input
\emph{and} accepts only in the initial node:
a reversible automaton with $4nk+1$ states,
described in Section~\ref{section_upper_bounds}, will do.

This section establishes a close lower bound on this transformation. 
The witness $n$-state automaton is the same as
in Sections~\ref{section_returning}--\ref{section_halting},
for which Theorem~\ref{theorem_ret} asserts that
a returning automaton needs at least $2(n-1)(k-3)$ states,
whereas Theorem~\ref{theorem_stop} proves that
a halting automaton needs at least $2(n-1)(k-3)$ states.
The goal is to prove that these two sets of states must be disjoint,
leading to the following lower bound.

\begin{theorem} \label{theorem_ret_stop}
For every $k \geqslant 4$,
there exists a signature with $k$ directions,
such that for every $n \geqslant 2$,
there is an $n$-state graph-walking automaton,
such that every returning and halting automaton recognizing the same set of graphs
must have at least $4(n-1)(k-3)$ states.
\end{theorem}

As before, the automaton is obtained from $A$ by Lemma~\ref{lemma_diod_forward}.
For the argument to proceed,
the following property needs to be established.

\begin{lemma}[cf.~Lemma~\ref{returning_automaton_reaches_n_minus_1_states_by_minus_a_lemma}]
\label{returning_and_stopping_automaton_reaches_2n_minus_2_states_by_minus_a_lemma}
Every returning and halting automaton
that recognizes the same set of graphs as $A$,
and has at most $4nk$ states,
enters at least $2(n-1)$ distinct states
after transitions in the direction $-a$.
\end{lemma}

Consider any such returning and halting automaton.
Since it is returning,
as shown in Lemma~\ref{returning_automaton_reaches_n_minus_1_states_by_minus_a_lemma},
on the graph $G_{n,k}^{accept}$,
the automaton uses a periodic sequence of states to return from $(M+8nk,1)$ to $v_0$.
Since it is at the same time halting,
Lemma~\ref{stopping_automaton_reaches_n_minus_1_states_by_minus_a_lemma}
asserts that on the graph $G$
it uses another periodic sequence of states
to escape the cycle after visiting $v_{joint}$.
Each of these two sequences makes transitions in the direction $-a$ in at least $n-1$ distinct states.
It remains to prove that these sequences are disjoint.

Suppose the sequences have a common element,
then they coincide up to a cyclic shift.
Then it is possible to modify $G$
so that the computation coming to $v_{joint}$
later continued as the computation on $G_{n,k}^{accept}$,
and led to acceptance.

\begin{proof}[Proof of Lemma~\ref{returning_and_stopping_automaton_reaches_2n_minus_2_states_by_minus_a_lemma}.]
Let $A'$ be any such returning and halting automaton.
Since it is returning,
on the graph $G_{n,k}^{accept}$,
it must come to the node $(M+8nk,1)$ in order to see the label $c_{acc}$,
and then find its way back to $v_0$.
Consider the case when the automaton's last visit
to any of the ends of the upper chain
is to $(M+8nk, 1)$
(the other case is proved similarly).

On the way from $(M+8nk, 1)$ to $v_0$,
the automaton sees only labels $c$.
At some point not far from $(M+8nk, 1)$, it starts behaving periodically,
repeating a certain sequence of states $q_1, \ldots, q_p$, with $p \leqslant 4nk$,
and moving in a sequence of directions $d_1, \ldots, d_p$.
The sequence $d_1, \ldots, d_p$ should move the automaton to the left,
so it contains more occurrences of $-a$ than of $a$.
Let $v_{\text{departure}}$ be the node at which the periodic behaviour starts,
visited in the state $q_1$.
Lemma~\ref{backwards_traversal_requires_n_minus_1_by_a_and_n_minus_1_by_minus_a_lemma}
asserts that the direction $-a$ occurs in the sequence $d_1, \ldots, d_p$
at least $n-1$ times.

Since $A'$ is halting,
the arguments in Lemma~\ref{stopping_automaton_reaches_n_minus_1_states_by_minus_a_lemma} also apply.
The lemma used a graph $G$ with a node $v_{joint}$
that merges $(M+8nk,1)$ and $(-8nk,1)$.
The automaton $A'$ must visit $v_{joint}$
in order to tell $G$ from $G_{n,k}^{accept}$.
Because the node $v_{joint}$ is far from any nodes labelled not with $c$,
when the automaton first comes to $v_{joint}$,
it repeats periodic sequences of states $q'_1, \ldots, q'_t$
and directions $d'_1, \ldots, d'_t$,
with $t \leqslant 4nk$.
Since this periodic behaviour must eventually stop,
this sequence must lead $A'$ from $v_{joint}$ to one of the ends of the lower chain.
Then, by Lemma~\ref{backwards_traversal_requires_n_minus_1_by_a_and_n_minus_1_by_minus_a_lemma},
there are at least $n-1$ occurrences of $-a$
in the sequence $d'_1, \ldots, d'_t$.

It remains to prove that none of the states $q_1, \ldots q_p$
may coincide with any of the states $q'_1, \ldots, q'_t$.
Suppose the contrary, that the periodic sequences are not disjoint.
Then both cycles consist of the same states,
cyclically shifted, and $p=t$.
The goal is to construct a graph without any labels $c_{acc}$,
which, however, would be accepted by $A'$, leading to a contradiction.

For each $x \in \{0, \ldots, 4nk\}$,
consider a graph $G_x$,
which is obtained from the graph $G$
by prolonging the upper cycle with $x$ extra nodes, all labelled with $c$.
Then the graph $G$ from Lemma~\ref{stopping_automaton_reaches_n_minus_1_states_by_minus_a_lemma}
is the graph $G_0$.

The automaton is known to accept the graph $G_{n,k}^{accept}$
starting from the node $v_{departure}$ in the state $q_1$,
without visiting either end of the upper chain.
In each graph $G_x$, let $v_{departure}$ be the node
located at the same distance from the nearest bridge to the lower chain.
If, on the graph $G_x$,
the automaton can be ``lured'' into this node in the state $q_1$,
then this graph will be accepted.

Since the sequences of directions $d_1, \ldots, d_p$
and $d_1', \ldots, d_t'$
coincide up to cyclic shift,
the direction $-a$ occurs more often than $a$ in $d_1', \ldots, d_t'$.
Then, when the automaton working on $G$
visits the node $v_{joint}$ for the first time,
it is moving in the direction $-a$,
and accordingly visits the node $v_{departure}$ after visiting $v_{joint}$.

Then it is possible to choose the length $x$
of a sequence of edges inserted after $v_{joint}$,
so that the automaton comes to the node $v_{departure}$ in the state $q_1$.
Then the automaton $A'$ accepts the graph $G_x$, and this is impossible.

A contradiction has thus been obtained,
and hence the sequences of states
$q_1, \ldots, q_p$ and $q_1', \ldots, q_t'$ are disjoint.
Together, they contain at least $2(n-1)$ distinct states
that the automaton enters after transitions in the direction $-a$.
\end{proof}

The proof of the theorem is inferred
from Lemmata~\ref{lemma_diod_forward}, \ref{New_THE_LEMMA}, \ref{lemma_diode_backward}
and~\ref{returning_and_stopping_automaton_reaches_2n_minus_2_states_by_minus_a_lemma},
as in the earlier arguments.

\section{Lower bound on the size of reversible automata}\label{section_reversible}

For the transformation of a GWA with $n$ states and $k$ directions
to a reversible automaton,
$4nk+1$ states are sufficient.
A close lower bound shall now be established.

\begin{theorem} \label{theorem_reversible}
For every $k \geqslant 4$,
there exists a signature with $k$ directions,
such that for every $n \geqslant 2$,
there is an $n$-state GWA,
such that 
every reversible GWA recognizing the same set of graphs
has at least $4(n-1)(k-3)-1$ states.
\end{theorem}
\begin{proof}
By Theorem~\ref{theorem_ret_stop}, there is such an $n$-state automaton $A'$
that every returning and halting automaton
recognizing the same set of graphs
has at least $4(n-1)(k-3)$ states.
Suppose that there is a reversible automaton
with fewer than $4(n-1)(k-3)-1$ states
that accepts the same graphs as $A'$.
Let $m$ be the number of states in it.
Then, by the construction of reversing a reversible automaton
given by Kunc and Okhotin~\cite{KuncOkhotin_reversible},
there is a returning and halting automaton with $m+1$ states,
that is, with fewer than $4(n-1)(k-3)$ states.
This contradicts Theorem~\ref{theorem_ret_stop}.
\end{proof}

\section{Conclusion}

The new bounds on the complexity of transforming graph-walking automata
to automata with returning, halting and reversibility properties
are fairly tight. 
However, for their important special cases, such as two-way finite automata (2DFA)
and tree-walking automata (TWA),
the gaps between lower bounds and upper bounds are still substantial.

For an $n$-state 2DFA,
the upper bound for making it halting is $4n+\mathrm{const}$ states~\cite{GeffertMereghettiPighizzini}.
No lower bound is known,
and any lower bound would be interesting to obtain.
A 2DFA can be made reversible using $4n+3$ states~\cite{KuncOkhotin_reversible},
with a lower bound of $2n-2$ states~\cite{rev2dfa};
it would be interesting to improve these bounds.

The same question applies to tree-walking automata:
they can be made halting~\cite{MuschollSamuelidesSegoufin},
and, for $k$-ary trees,
it is sufficient to use $4kn+2k+1$ states
to obtain a reversible automaton~\cite{KuncOkhotin_reversible}.
No lower bounds are known, and this subject is suggested for further research.

Furthermore, it would be interesting to try to apply the lower bound methods for GWA
to limited memory algorithms for navigation in graphs.


\begin{thebibliography}{99}

\bibitem{BojanczykColcombet_det} M. Boja\'nczyk, T. Colcombet,
	\href{http://dx.doi.org/10.1016/j.tcs.2005.10.031}
	{``Tree-walking automata cannot be determinized''},
	\emph{Theoretical Computer Science},
	350:2--3 (2006), 164--173.

\bibitem{Budach} L. Budach,
	\href{http://dx.doi.org/10.1002/mana.19780860120}
	{``Automata and labyrinths''},
	\emph{Mathematische Nachrichten},
	86:1 (1978), 195--282.

\bibitem{DisserHackfeldKlimm} Y. Disser, J. Hackfeld, M. Klimm,
	\href{https://doi.org/10.1137/1.9781611974331.ch3}
	{``Undirected graph exploration with $O(\log \log n)$ pebbles''},
	\emph{Proceedings of the Twenty-Seventh Annual ACM-SIAM Symposium on Discrete Algorithms}
	(SODA 2016, Arlington, USA, 10--12 January 2016),
	25--39.

\bibitem{ElmasryHagerupKammer} A. Elmasry, T. Hagerup, F. Kammer,
	\href{https://doi.org/10.4230/LIPIcs.STACS.2015.288}
	{``Space-efficient basic graph algorithms''},
	\emph{STACS 2015},
	288--301.

\bibitem{FraigniaudIlcinkasPeerPelcPeleg} P. Fraigniaud, D. Ilcinkas, G. Peer, A. Pelc, D. Peleg,
	\href{http://dx.doi.org/10.1016/j.tcs.2005.07.014}
	{``Graph exploration by a finite automaton''},
	\emph{Theoretical Computer Science},
	345:2--3 (2005), 331--344.

\bibitem{GeffertMereghettiPighizzini} V. Geffert, C. Mereghetti, G. Pighizzini,
	\href{http://dx.doi.org/10.1016/j.ic.2007.01.008}
	{``Complementing two-way finite automata''},
	\emph{Information and Computation},
	205:8 (2007), 1173--1187.

\bibitem{KondacsWatrous} A. Kondacs, J. Watrous,
	\href{http:dx.doi.org/10.1109/SFCS.1997.646094}
	{``On the power of quantum finite state automata''},
	\emph{38th Annual Symposium on Foundations of Computer Science}
	(FOCS 1997, Miami Beach, Florida, USA, 19--22 October 1997),
	IEEE, 66--75.

\bibitem{rev2dfa} M. Kunc, A. Okhotin,
	\href{https://tucs.fi/publications/view/?pub_id=tOkKu11a}
	{``Reversible two-way finite automata over a unary alphabet''},
	TUCS Technical Report 1024,
	Turku Centre for Computer Science,
	December 2011.

\bibitem{KuncOkhotin_reversible} M. Kunc, A. Okhotin,
	\href{https://doi.org/10.1016/j.ic.2020.104631}
	{``Reversibility of computations in graph-walking automata''},
	\emph{Information and Computation},
	275 (2020), article 104631.

\bibitem{Landauer} R. Landauer,
	\href{http://dx.doi.org/10.1147/rd.53.0183}
	{``Irreversibility and heat generation in the computing process''},
	\emph{IBM Journal of Research and Development},
	5:3 (1961), 183--191.

\bibitem{LangeMcKenzieTapp} K.-J. Lange, P. McKenzie, A. Tapp,
	\href{http://dx.doi.org/10.1006/jcss.1999.1672}
	{``Reversible space equals deterministic space''},
	\emph{Journal of Computer and System Sciences},
	60:2 (2000), 354--367.

\bibitem{Morita} K. Morita,
	\href{http://dx.doi.org/10.1007/978-3-642-36315-3_3}
	{``A deterministic two-way multi-head finite automaton can be converted into a reversible one with the same number of heads''},
	\emph{Reversible Computation} 
	(RC 2012, Copenhagen, Denmark, 2--3 July 2012),
	LNCS 7581,
	29--43.

\bibitem{MuschollSamuelidesSegoufin} A. Muscholl, M. Samuelides, L. Segoufin,
	\href{http://dx.doi.org/10.1016/j.ipl.2005.09.017}
	{``Complementing deterministic tree-walking automata''},
	\emph{Information Processing Letters},
	99:1 (2006), 33-39.

\bibitem{Sipser_halting} M. Sipser,
	\href{http://dx.doi.org/10.1016/0304-3975(80)90053-5}
	{``Halting space-bounded computations''},
	\emph{Theoretical Computer Science},
	10:3 (1980), 335--338.

\end{thebibliography}
\end{document}